\documentclass[twocolumn,english,aps,pra,reprint,nofootinbib,superscriptaddress]{revtex4-1}
\usepackage[T1]{fontenc}
\usepackage[latin9]{inputenc}
\usepackage{amsmath}
\usepackage{amssymb}
\usepackage{graphicx}
\usepackage{braket}
\usepackage{comment}

\usepackage{color}
\usepackage{float}
\usepackage{hyperref}
\usepackage{makecell}

\hypersetup{
    colorlinks,
    linkcolor={black},
    citecolor={black},
    urlcolor={blue}
}

\makeatletter
\@ifundefined{textcolor}{}
{%
 \definecolor{BLACK}{gray}{0}
 \definecolor{WHITE}{gray}{1}
 \definecolor{RED}{rgb}{1,0,0}
 \definecolor{GREEN}{rgb}{0,1,0}
 \definecolor{BLUE}{rgb}{0,0,1}
 \definecolor{CYAN}{cmyk}{1,0,0,0}
 \definecolor{MAGENTA}{cmyk}{0,1,0,0}
 \definecolor{YELLOW}{cmyk}{0,0,1,0}
}

\usepackage{pslatex}

\usepackage{babel}

\makeatother

\begin{document}

\title{Establishing a nearly closed cycling transition in a polyatomic molecule
}

\author{Louis Baum}
\email{louisbaum@g.harvard.edu}
\affiliation{Harvard-MIT Center for Ultracold Atoms, Cambridge, MA 02138, USA}
\affiliation{Department of Physics, Harvard University, Cambridge, MA 02138, USA}

\author{Nathaniel B. Vilas}
\affiliation{Harvard-MIT Center for Ultracold Atoms, Cambridge, MA 02138, USA}
\affiliation{Department of Physics, Harvard University, Cambridge, MA 02138, USA}

\author{Christian Hallas}
\affiliation{Harvard-MIT Center for Ultracold Atoms, Cambridge, MA 02138, USA}
\affiliation{Department of Physics, Harvard University, Cambridge, MA 02138, USA}

\author{\\Benjamin L. Augenbraun}
\affiliation{Harvard-MIT Center for Ultracold Atoms, Cambridge, MA 02138, USA}
\affiliation{Department of Physics, Harvard University, Cambridge, MA 02138, USA}

\author{Shivam Raval}
\affiliation{Harvard-MIT Center for Ultracold Atoms, Cambridge, MA 02138, USA}
\affiliation{Department of Physics, Harvard University, Cambridge, MA 02138, USA}

\author{Debayan Mitra}
\affiliation{Harvard-MIT Center for Ultracold Atoms, Cambridge, MA 02138, USA}
\affiliation{Department of Physics, Harvard University, Cambridge, MA 02138, USA}

\author{John M. Doyle}
\affiliation{Harvard-MIT Center for Ultracold Atoms, Cambridge, MA 02138, USA}
\affiliation{Department of Physics, Harvard University, Cambridge, MA 02138, USA}

\date{\today}

\begin{abstract}

  We study optical cycling in the polar free radical calcium monohydroxide (CaOH) and establish an experimental path towards scattering $\sim$$10^4$ photons. We report rovibronic branching ratio measurements with precision at the $\sim10^{-4}$ level and observe weak symmetry-forbidden decays to bending modes with non-zero vibrational angular momentum. Calculations are in excellent agreement with these measurements and predict additional decay pathways. Additionally, we perform high-resolution spectroscopy of the $\widetilde{\text{X}}\,^2\Sigma^+(12^00)$ and $\widetilde{\text{X}}\,^2\Sigma^+(12^20)$ hybrid vibrational states of CaOH. These advances establish a path towards radiative slowing, 3D magneto-optical trapping, and sub-Doppler cooling of CaOH. 

\end{abstract}
\maketitle

\section{Introduction}

Laser cooling, one of the cornerstones of atomic, molecular and optical physics~\cite{chu1998nobel,phillips1998nobel}, has enabled wide-ranging scientific applications including ultra-precise clocks~\cite{Swallows1043}, quantum simulation of many body systems~\cite{bloch2008many}, and novel quantum computation platforms~\cite{endres2016atom,bernien2017probing,anderegg2019optical}. Extension of laser cooling techniques to polyatomic molecules is at the forefront of efforts to produce ultracold samples of polyatomic species. Ultracold polyatomic molecules have been proposed for a wide range of multidisciplinary applications, including quantum simulation~\cite{wall2013simulating,wall2015realizing} and computation~\cite{yu2019scalable}, quantum chemistry and collisions~\cite{augustovicova2019collisions, bohn2017cold}, and tests of fundamental physics, including searches for the electron electric dipole moment (eEDM)~\cite{kozyryev2017precision}, ultralight dark matter~\cite{kozyryev2018enhanced}, and fundamental parity violation~\cite{norrgardNuclear2019}.

In the past several years, enormous strides have been made in direct laser cooling of polyatomic molecules, with molecular beams of SrOH~\cite{kozyryev2017sisyphus, kozyryev2018coherent}, YbOH~\cite{augenbraun2019laser}, CaOH~\cite{baum20201d}, and CaOCH$_3$~\cite{mitra2020direct} all cooled near or below 1~mK in one transverse dimension. The ability to rapidly scatter a large number of photons is at the heart of these cooling techniques, which typically require an estimated 10$^4$ scattered photons for successful confinement in a magneto-optical trap (MOT). Thus far, experimental efforts have successfully enabled scattering up to $\sim$10$^3$ photons in polyatomic molecules~\cite{baum20201d,augenbraun2019laser}. 

To establish a path to a MOT, vibrational branching ratios (VBRs)  need to be determined with accuracy at, or exceeding, the 10$^{-4}$ level. Measurements of VBRs have previously been performed in polyatomic molecules using dispersed laser fluorescence~\cite{kozyryev2019determination,Mengesha2020Branching,nguyen2018fluorescence}. However, previously reported measurements of CaOH using this technique~\cite{kozyryev2019determination} do not provide sufficient sensitivity to identify the loss channels needed to scatter 10$^4$ photons. 

In this work, we describe a measurement of VBRs in CaOH by detecting accumulated population in excited rotational and vibrational levels after hundreds of photons are scattered. Cycling multiple photons enhances the measurement sensitivity, probing VBRs at the 10$^{-4}$ level. Several weak, symmetry-forbidden decays are observed. We describe and benchmark calculations of branching ratios that include vibronic perturbations not considered in previous work. Using these calculations, we propose a photon cycling scheme with the predicted capability to scatter an average of $\sim$10$^4$ photons per molecule. Finally, we perform high-resolution spectroscopy of the $\widetilde{\text{X}}\,^2\Sigma^+(12^00)$ and $\widetilde{\text{X}}\,^2\Sigma^+(12^20)$ hybrid vibrational modes. It is expected that repumping one or both of these states will be necessary to enable radiative slowing, trapping, and sub-Doppler cooling of CaOH.

\section{Vibrational Branching Overview and Measurements}
\label{VBRmeasurements}

In order to create a sustained cycling transition in CaOH, it is necessary to close loss channels due to both vibration and rotation. The relative probability of spontaneous decay to different vibrational states is described by vibrational branching ratios (VBRs). These are closely related to the Franck-Condon factors (FCFs) of the molecule, defined by the overlap integral between vibrational wavefunctions in the ground and excited states (see section \ref{Cals}). CaOH is an example of a broad class of polyatomic molecules that have been identified as amenable to laser cooling due to strong electronic transitions and near-diagonal FCFs~\cite{kozyryev2016proposal,kozyryev2019determination,augenbraun2020molecular}.

\begin{figure}
    \centering
    \includegraphics[width = \linewidth]{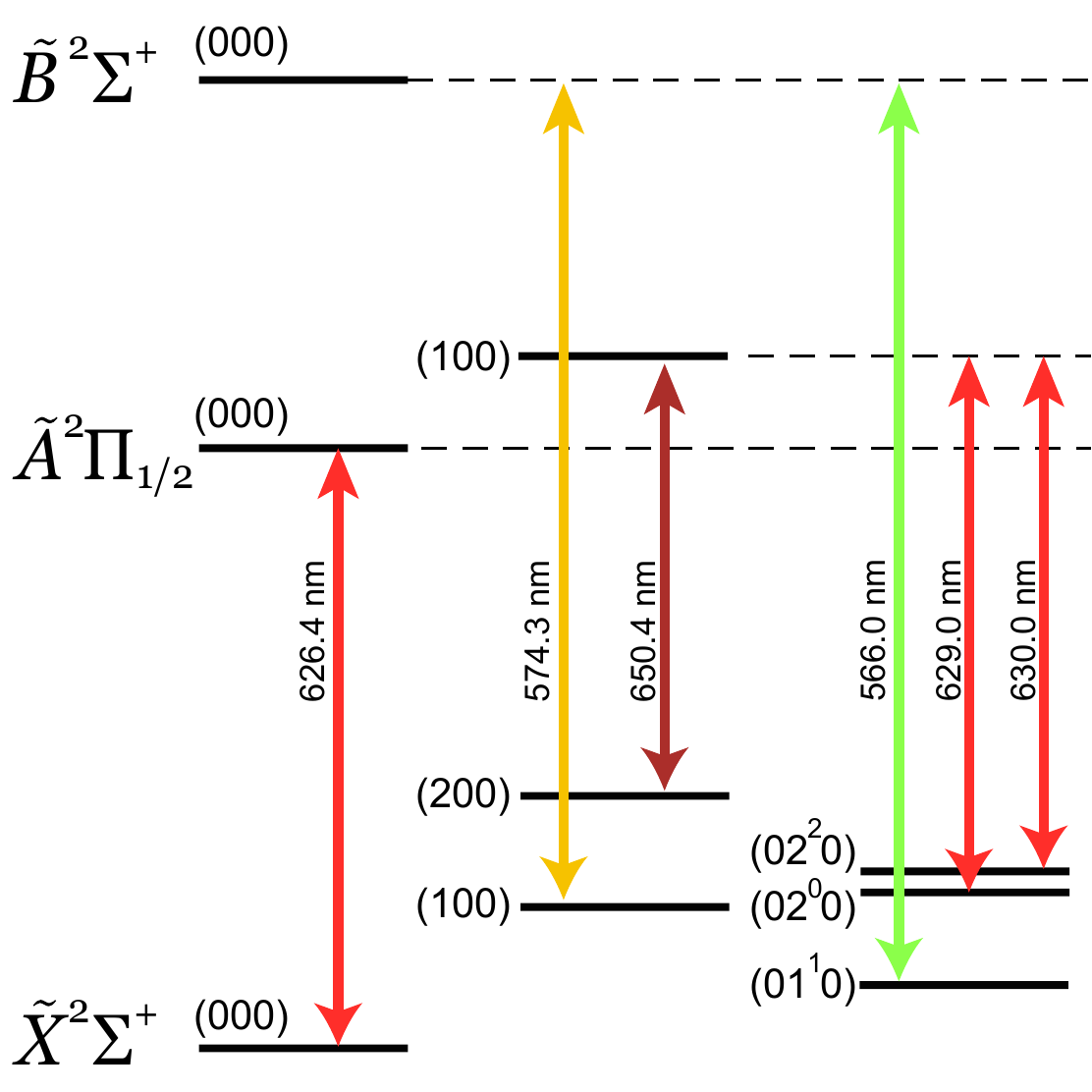}
    \caption{Vibrational structure of CaOH and laser cooling scheme experimentally considered in this work.}
    \label{fig:cyclingscheme}
\end{figure}

The main laser cooling transition in CaOH is the $\widetilde{\text{X}}^2\Sigma^+\left(000\right) \rightarrow\widetilde{\text{A}}^2\Pi_{1/2}\left(000\right)$ transition. Vibrational states are labeled with the quantum numbers $\left(v_1,{v_2}^\ell,v_3\right)$, where $v_1$, $v_2$, and $v_3$ are the number of quanta in the symmetric (predominantly Ca--O) stretching, bending, and antisymmetric (predominantly O--H) stretching modes, respectively. $\ell$ labels the nuclear orbital angular momentum in the bending mode~\cite{herzberg1966molecular}.  The highly diagonal FCFs of the $\widetilde{\text{A}}^2\Pi_{1/2}\left(000\right)$ state suppress spontaneous decay to excited vibrational states during each scattering event; nonetheless, some excited vibrational states must be repumped due to significant optical pumping after many photons are scattered. In this work, we experimentally characterize the photon cycling scheme depicted in Fig. \ref{fig:cyclingscheme} and show that it is capable of scattering $\sim$1000 photons. Note that multiple excited electronic states are used to cycle photons, which must be taken into account when comparing measured and calculated VBRs as discussed in section \ref{Cals}.

\begin{figure}
    \centering
    \includegraphics[width = \linewidth]{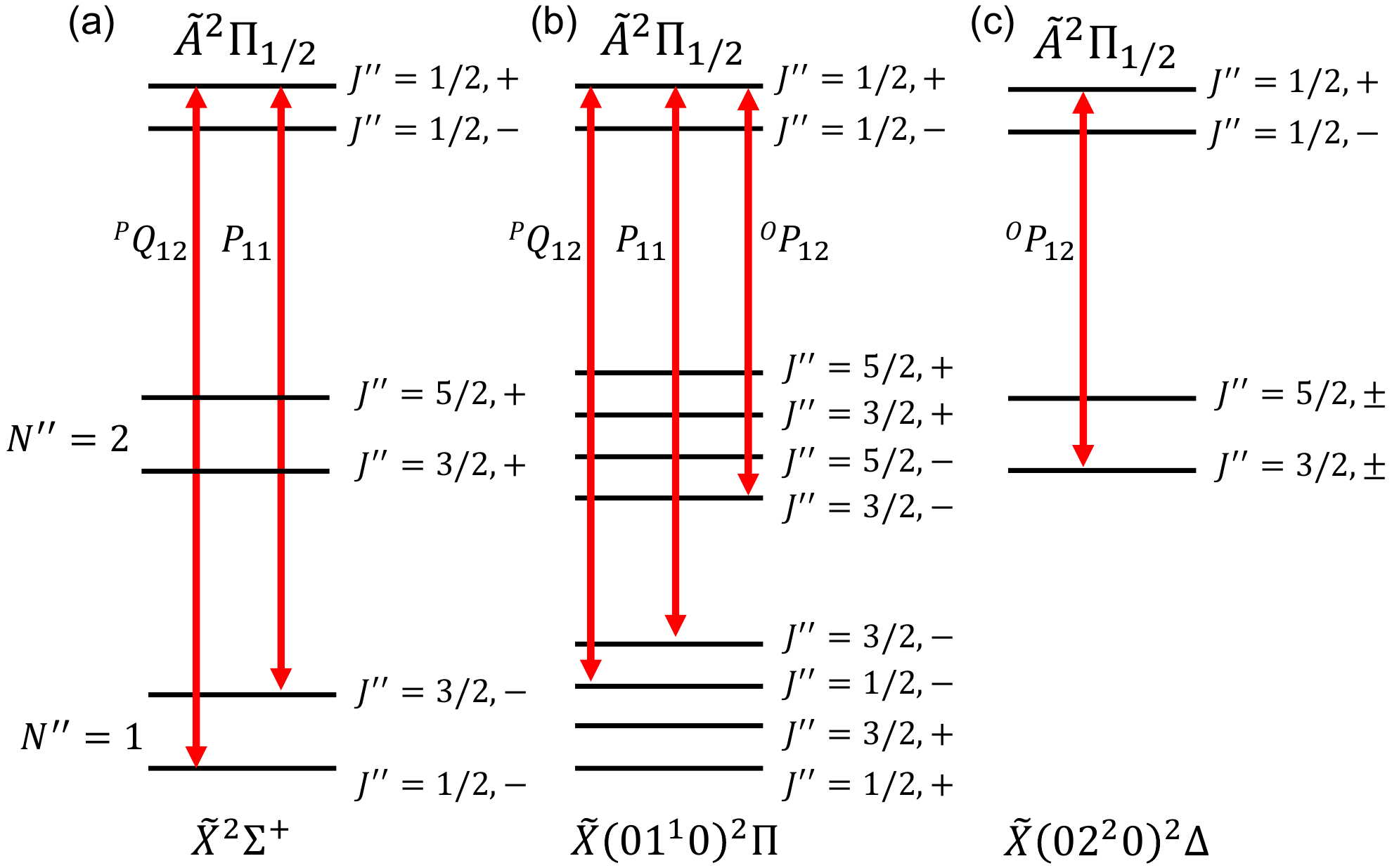}
    \caption{Allowed rotational transitions for $\widetilde{\text{A}}\,^2\Pi_{1/2} \leftarrow \widetilde{\text{X}}\,^2\Sigma^+$ CaOH laser cooling. (a) All non-degenerate ground states of $^2\Sigma^+$ symmetry (e.g., $(000)$, $(100)$, $(02^00)$) have only two allowed rotational components split by $\sim$50 MHz. Hyperfine splittings of 1.5 MHz and 7 kHz in the $J''=\frac{3}{2}$ and $J''=\frac{1}{2}$ states, respectively~\cite{scurlock1993hyperfine}, are not shown. (b) Because of parity doubling in states of $^2\Pi$ symmetry (e.g., $(01^10)$), additional decay is allowed to the $N''=2$ rotational level. This must be repumped using an additional laser frequency $\sim$40 GHz away from the $N''=1$ transition. (c) States with $^2\Delta$ symmetry (e.g., $(02^20)$, $(12^20)$) have no $N''=1$ component, so only the $^OP_{12}$ line must be addressed. Note that decay from $\widetilde{A}^2\Pi_{1/2}(000)$ to the $\ell\neq 0$ states in (b)-(c) is symmetry-forbidden, and only occurs due to  vibronic perturbations in the excited state (Sec. \ref{Cals}).}
    \label{fig:rotational}
\end{figure}

For states with $\ell = 0$,  rotational losses are prevented by driving the $P_1$ ($N''=1$) and $^PQ_{12}$ ($N''=1$) transitions~\cite{di2004laser}. In the ground electronic state of CaOH ($\widetilde{\text{X}}\,^2\Sigma^+$), the $N''=1$ manifold is split by the spin-rotation interaction into $J''=1/2$ and $J''=3/2$ components separated by 52 MHz. Both components are addressed by using an acousto-optic modulator (AOM) to add a frequency-shifted sideband to the laser light. All hyperfine splittings are below the natural linewidth of the optical transition (Fig. \ref{fig:rotational}(a)). For states with $\ell \neq 0$, parity doubling additionally enables decay to $N''=2$, which must be addressed to achieve full rotational closure (Fig. \ref{fig:rotational}(b,c)).

\begin{figure*}
    \centering
    \includegraphics[width=.8\textwidth]{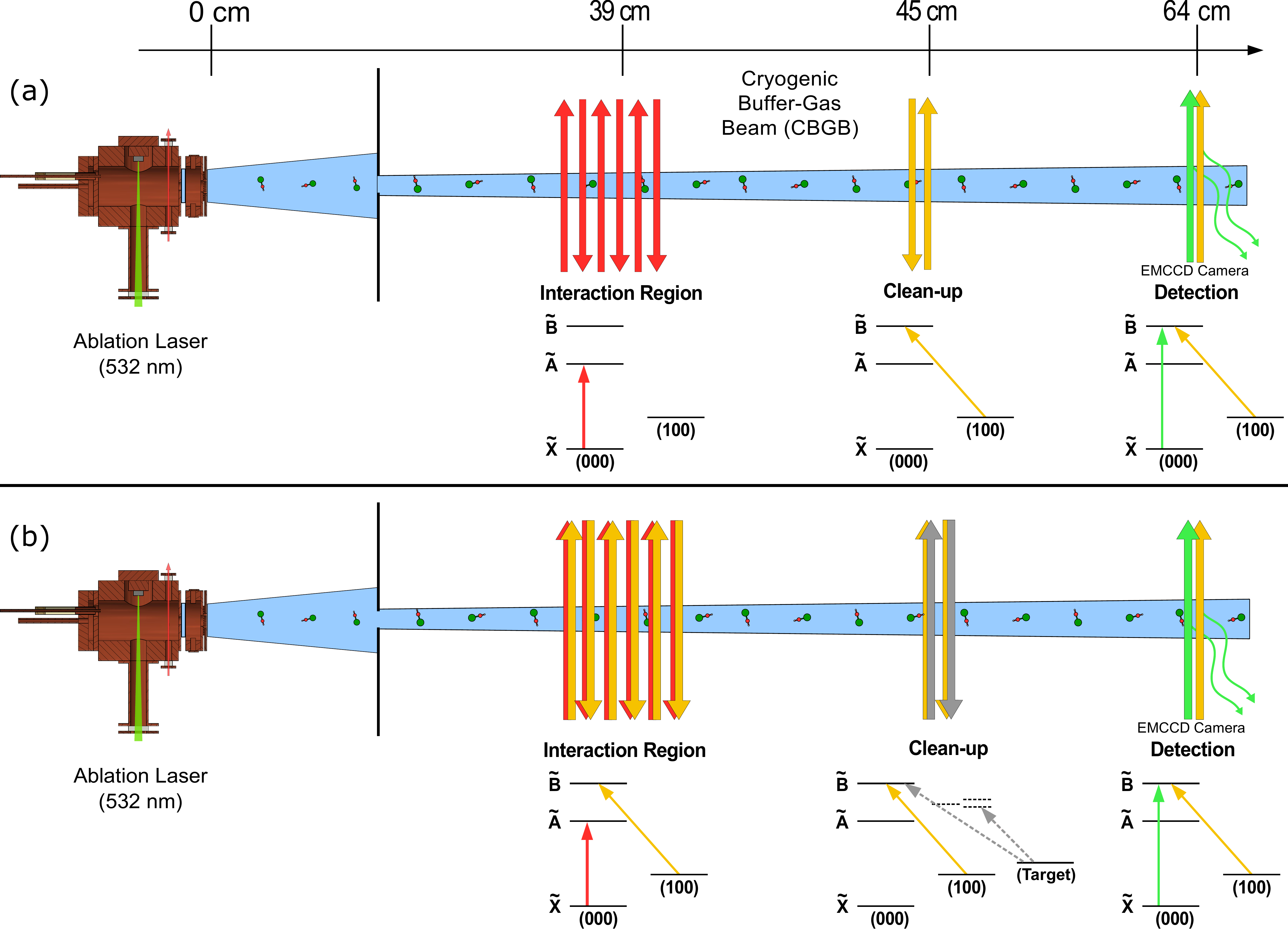}
    \caption[VBR Experimental configuration]{(a) Experimental configuration used to measure vibrational branching ratios (VBRs) from $\widetilde{\text{A}}^2\Pi_{1/2}\left(000\right)$ to $\widetilde{\text{X}}^2\Sigma^+\left(000\right)$ and $\left(100\right)$. (b) Experimental configuration used to measure VBRs to higher vibrational states. Colors indicate which transitions are addressed in each optically accessible region. Grey and dotted lines indicate that the specific transition used depends on the vibrational state targeted in a particular measurement.}
    \label{fig:experimentaldiagram}
\end{figure*}

CaOH molecules are produced using a cryogenic buffer gas source described in previous work~\cite{baum20201d} and depicted in Fig. \ref{fig:experimentaldiagram}. Densities of $\sim$ 10$^{10}$ cm$^{-3}$ in a single rotational state ($N'' = 1$) are routinely achieved. CaOH molecules are extracted from a two-stage buffer gas cell and form a cryogenic buffer-gas beam (CBGB)~\cite{hutzler2012buffer}. The CBGB is collimated by a 3$\times$3 mm aperture to ensure that all molecules are addressed with the applied laser light. 

We use a combination of two measurements to determine the VBRs of CaOH. First, we determine VBRs from $\widetilde{\text{A}}^2\Pi_{1/2}(000)$ to the $\widetilde{\text{X}}^2\Sigma^+\left(000\right)$ and $\widetilde{\text{X}}^2\Sigma^+\left(100\right)$ states by measuring relative population accumulation in the $(100)$ level after optically pumping molecules out of the $(000)$ ground state. This measurement is combined with previously published results \cite{kozyryev2019determination} to more precisely quantify the vibrational branching ratios.   Secondly, we determine VBRs to \textit{other} excited vibrational states in the $\widetilde{\text{X}}$ manifold by optically pumping molecules out of $\widetilde{\text{X}}^2\Sigma^+\left(000\right)$ and $\widetilde{\text{X}}^2\Sigma^+\left(100\right)$ and then measuring the population increase in these other vibrational levels. Relative measurements of recovered populations, combined with measurements of the total population lost to unaddressed levels, can be used to reconstruct the vibrational branching ratios of this laser cooling scheme.

\subsection{Vibrational Branching Ratios to $\widetilde{\text{X}}$(000) and (100)}

\begin{figure}
     \centering
     \includegraphics[width = \linewidth]{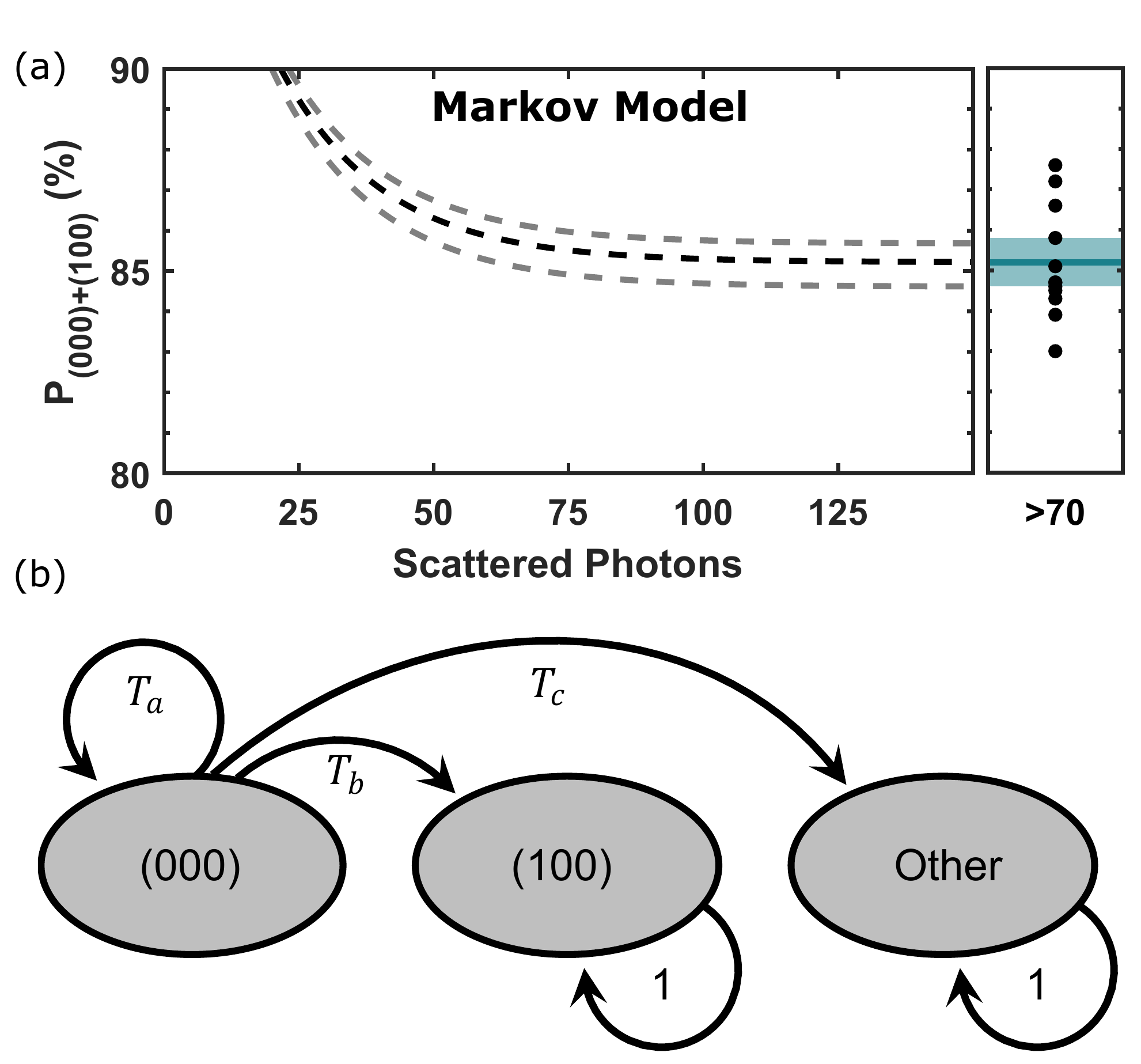}
     \caption{\label{fig:100Markov}(a) Recovered population as a function of scattered photons given by the discrete Markov chain model. The black line depicts the molecular population in both $\widetilde{\text{X}}(000)$ and $\widetilde{\text{X}}(100)$ vibrational states. The 1 $\sigma$ uncertainly in the ratio of $T_a/T_b$ is indicated by the grey dashed lines. Experimental measurements are plotted separately as the number of scattered photons is poorly defined; however, depletion measurements confirm that these data points are within the asymptotic limit ( > 70 scattered photons). The 12 black points indicate individual measurements, each averaging over 75 experimental cycles. The scatter in these measurements is due primarily to fluctuations in background level caused by scattered light. The mean of these experimental measurements is given by the blue line, where the shaded blue region denotes the standard error in the mean. The results indicate that the combined VBR to both the $\widetilde{\text{X}}(000)$ and $\widetilde{\text{X}}(100)$ states in this laser cooling scheme is 0.9918(8).
     (b)  Diagrammatic representation of the Markov chain model. Population evolves between three states representing $\widetilde{\text{X}}(000)$,  $\widetilde{\text{X}}(100)$, and all other vibrational states. For each spontaneous emission event the molecular population evolves as shown by the black arrows with the probabilities listed. For this measurement the interaction lasers only address the (000) state. Both (100) and ``Other'' are absorbing states as indicated by the unit probability of remaining in the same state. Spontaneous vibrational decay is negligible over the experimental timescale. }
     
 \end{figure}

In the first measurement, the $\widetilde{\text{X}}^2\Sigma^+\left(000\right) \rightarrow\widetilde{\text{A}}^2\Pi_{1/2}\left(000\right)$ laser is retroreflected through the interaction region as depicted in Fig. \ref{fig:experimentaldiagram}(a). Molecules interact with this light field, cycling photons until >95\% of the molecular population has decayed to a dark vibrational state. The increase in molecular population in each excited vibrational state is directly proportional to the vibrational branching ratio to that state. A laser addressing the $\widetilde{\text{X}}^2\Sigma^+\left(100\right) \rightarrow\widetilde{\text{B}}^2\Sigma^+\left(000\right)$ transition recovers population lost to the $\widetilde{\text{X}}^2\Sigma^+(100)$ state. Molecules are detected with lasers addressing the $\widetilde{\text{X}}^2\Sigma^+\left(000\right) \rightarrow\widetilde{\text{B}}^2\Sigma^+\left(000\right)$ and $\widetilde{\text{X}}^2\Sigma^+\left(100\right) \rightarrow\widetilde{\text{B}}^2\Sigma^+\left(000\right)$ transitions, and the resulting laser-induced fluorescence is imaged onto an electron multiplying charged coupled device (EMCCD) camera. Absolute population drifts due to ablation fluctuations are mitigated by normalizing the recovered population with interleaved measurements of the total population taken with the light in the interaction region blocked. The results of this measurement are presented in Fig. \ref{fig:100Markov}(a), where the fractional population recovered is $P_{rec}$ = 85.2(6)\%.  

We model this process as a discrete Markov chain with two absorbing states as indicated in Fig. \ref{fig:100Markov}(b). Each discrete step corresponds to a spontaneous emission event which causes the population to evolve according to the expression

\begin{equation}
    \begin{bmatrix} P^{\,n+1}_{(000)} \\P^{\,n+1}_{(100)} \\ P^{\,n+1}_\text{other}  \end{bmatrix} = \begin{bmatrix} T_a& 0 & 0 \\ T_b & 1 & 0 \\ T_c & 0 & 1 \end{bmatrix} \begin{bmatrix}  P^{\,n}_{(000)} \\P^{\,n}_{(100)} \\ P^{\,n}_\text{other}   \end{bmatrix}.
\end{equation}

In this expression the parameters $T_a$, $T_b$, and $T_c$ correspond to the vibrational branching ratios from $\widetilde{\text{A}}(000)$ to $\widetilde{\text{X}}(000)$, $\widetilde{\text{X}}(100)$, and all other loss channels, respectively.  $P^n_{(000)}$, $P^n_{(100)}$, and $P^n_{\text{other}}$ are the molecular populations in $\widetilde{\text{X}}(000)$, $\widetilde{\text{X}}(100)$, and all other states, respectively, after $n$ spontaneous decay events. This formula can be recursively applied with the resulting population at any step ($n$) described by the expression

\begin{equation}
    \overrightarrow{P}_{n} = \begin{bmatrix} T_a& 0 & 0 \\ T_b & 1 & 0 \\T_c & 0 & 1 \end{bmatrix}^{n} \overrightarrow{P}_{0}.
\end{equation}

$\overrightarrow{P}_{0}$ is a vector that represents the initial distribution of molecular population and is experimentally determined. In this first measurement, we do not measure the final population $P^n_{\text{other}}$, and as a result our model is insensitive to this quantity. We represent $\overrightarrow{P}_{0}$ as

\begin{equation*}
    \overrightarrow{P}_{0} = \begin{bmatrix} P^0_{(000)} \\ P^0_{(100)} \\ P^0_{\text{other}} \end{bmatrix} = \begin{bmatrix} 0.92(1) \\ 0.08(1) \\ 0 \end{bmatrix}
\end{equation*}
where $P^0_{(000)}$ and $P^0_{(100)}$ are experimentally determined and $P^0_{\text{other}}$ is set to zero to reflect the insensitivity of the model. As photons are scattered, the percentage of molecular population in excited vibrational states converges to an asymptotic limit. Previous measurements of vibrational branching in Ref.~\cite{kozyryev2019determination} have measured the VBRs to $\widetilde{\text{X}}(000)$ and $\widetilde{\text{X}}(100)$: $T_a\,=\,0.957(2)$ and $T_b\,=\,0.043(2)$. However, these values sum to 1, which is nonphysical given the known decay to higher vibrational states. To account for this, we use these measurements to constrain the ratio $T_a / T_b$ =$\frac{0.957(2)}{0.043(2)}$ = 22.3(1.0). In addition, conservation of probability demands that $T_a\,+\,T_b\,+\,T_c\,=\,1$. These two constraints, combined with the experimental measurement of $P_{\text{rec}}$ described above, allow the ratio of $T_b / T_c$ to be uniquely determined.  The values of $T_a$, $T_b$, and $T_c$ are extracted from the model, and the combined vibrational branching ratio from $\widetilde{\text{A}}(000)$ to $\widetilde{\text{X}}(000)$ and $\widetilde{\text{X}}(100)$ is found to be $T_a\,+\,T_b$ = 0.9918(8). The uncertainty in this value is obtained by allowing each of $P_{rec}$ and $T_a/T_b$ to vary by up to 1 standard error in the model, computing the corresponding $T_a$ + $T_b$ value, and taking one half the maximum difference found.

\subsection{Vibrational Branching Ratios to $\widetilde{\text{X}}$(200), (02$^0$0), (02$^2$0), and (01$^1$0)}
\label{sec:HighVBR}

In order to measure the VBRs to higher-lying vibrational states, we deplete population from the $\widetilde{\text{X}}(000)$ and $\widetilde{\text{X}}(100)$ states and directly measure the accumulation of molecules in other excited vibrational levels. In the interaction region (see Fig. \ref{fig:experimentaldiagram}(b)), photon cycling lasers addressing the $\widetilde{\text{X}}^2\Sigma^+\left(000\right) \rightarrow\widetilde{\text{A}}^2\Pi_{1/2}\left(000\right)$ and $\widetilde{\text{X}}^2\Sigma^+\left(100\right) \rightarrow\widetilde{\text{B}}^2\Sigma^+\left(000\right)$ transitions are retroreflected between two mirrors for a total of $\sim$ 12 cm of interaction distance. This interaction length is sufficient to optically pump $\sim$ 90\% of the population to higher vibrational states, which corresponds to >250 scattered photons. Cycling multiple photons is crucial to this measurement as it allows for substantial optical pumping (10\% level population transfer) through small decay pathways. We selectively recover population from the states of interest by applying appropriate repumping lasers in the clean-up region.

The total loss probability must sum to 1 and the measured VBR to $\widetilde{\text{X}}(000)$ and $\widetilde{\text{X}}(100)$ accounts for 0.9918(8) of this total. The remaining decay probability to higher vibrational levels is assigned according to the fraction of population recovered from each state following the depletion and revival process. During these measurements we monitor both the total depletion from $\widetilde{\text{X}}(000)$ and $\widetilde{\text{X}}(100)$ as well as the natural population in the excited vibrational states of interest, both of which are accounted for in post-analysis as described in detail in Appendix \ref{app:ExpSeq}. The results are summarized in \autoref{tab:VBRComparison}. The measured VBRs are in agreement with previous results~\cite{kozyryev2019determination}. While addressing the five decay pathways listed in \autoref{tab:VBRComparison}, 13(1)\% of the depleted population is not recovered. We attribute this loss to yet higher-lying states and assign a corresponding VBR to all vibrational states not addressed in the laser cooling scheme in Fig. \ref{fig:cyclingscheme}.

\begin{table}
\begin{tabular}{c c c c c c c}
\Xhline{1pt}
\\[-1 em]
\Xhline{1pt} 
\\[-0.7 em]
Decay && Harmonic && Corrected\footnote{Corrections due to Renner-Teller and Fermi resonance couplings, as described in the text.} && Observed \\
\\[-1 em]
\Xcline{1-1}{1 pt}\Xcline{3-3}{1 pt}\Xcline{5-5}{1 pt}\Xcline{7-7}{1 pt}
\\[-.5 em]
$(000)$ && 0.957 && 0.955 && 0.9492(27) \\
$(100)$ && 0.042 && 0.038 && 0.0426(19) \\
$(200)$ && $0.6\times10^{-3}$ && $0.4\times 10^{-3}$ && $2.5(3)\times 10^{-3}$ \\
$(02^00)$ && $1.6\times 10^{-4}$ && $4.6\times 10^{-3}$ && $3.3(4)\times 10^{-3}$ \\
$(02^20)$ && 0 && $1.0\times 10^{-3}$ && $8.2(1.3)\times 10^{-4}$ \\
$(01^10),N''=1$ && 0 && $4.3\times 10^{-4}$ && $6.4(1.1)\times 10^{-4}$ \\
Other && $1.0\times 10^{-4}$ && $6.3\times 10^{-4}$ && $1.1(1)\times 10^{-3}$ \\
\\[-1 em]
\Xhline{1pt} 
\\[-1 em]
\Xhline{1pt} 
\\[-0.5 em]
\end{tabular}
\caption{Comparison of calculated and observed rovibronic branching ratios from the combined $\widetilde{\text{A}}\,^2\Pi_{1/2}(000)\leftarrow \widetilde{\text{X}}\,^2\Sigma^+(000) + \widetilde{\text{B}}\,^2\Sigma^+(000) \leftarrow \widetilde{\text{X}}\,^2\Sigma^+(100)$ photon cycling scheme used in this work. Branching ratio calculations based on unperturbed harmonic wavefunction overlap fail to adequately capture decay to the $\widetilde{\text{X}}\,^2\Sigma^+(02^00)$, $(02^20)$, and $(01^10)$ bending modes. These discrepancies are resolved by including Fermi resonance and Renner-Teller interactions. All measured and calculated branching ratios are rotationally resolved, with listed decays from $J'=1/2$, $(+)$ parity excited states to $\widetilde{\text{X}}\,^2\Sigma^+\left(v_1 v_2^\ell v_3\right)$, $N''=1$, $(-)$ parity ground states.}

\label{tab:VBRComparison}
\end{table}

\section{Calculated Branching Ratios}
\label{Cals}

The strength of generic, dipole-allowed rovibronic decays is governed by the Einstein A coefficient~\cite{bernath2005spectra}
\begin{equation}
   A_{J'\rightarrow J''} = \frac{16\pi^3 \nu^3 \lvert \langle \mu \rangle \rvert^2}{3\epsilon_0 h c^3(2J'+1)},
\label{eqn:EinsteinA}
\end{equation}
where $\lvert \langle \mu \rangle \rvert ^2 = \sum_{M'M''}|\langle \eta', v', J'M' \lvert \mu \rvert \eta'', v'', J'' M'' \rangle|^2$ is the degeneracy-weighted line strength of the electric dipole transition and $\nu$ is the transition frequency. The rovibronic states are $|\eta, v, JM\rangle$ and consist of electronic, $|\eta\rangle$, vibrational, $|v\rangle$, and rotational, $|JM\rangle$, components. The line strength may be approximately separated as follows:
\begin{equation}
    \lvert \langle \mu \rangle \rvert^2 \approx q_{v'-v''}|\mathbf{R}_e|^2 S^{J'}_{J''}(\zeta',\zeta''),
\end{equation}
where $|\mathbf{R}_e|^2$ is the electronic transition dipole moment and
\begin{equation}
    q_{v'-v''}  = |\braket{v''|v'}|^2
\label{eqn:FCFintegral}
\end{equation}
is the Franck-Condon factor (FCF). $S^{J'}_{J''}(\zeta', \zeta'')$ is the H\"onl-London (HL) factor, which characterizes the rotational line strength but depends in general on the vibronic states involved, labeled $|\zeta\rangle \equiv |\eta, v\rangle$ for notational convenience. Some relevant HL factors are given in Table \ref{tab:RotationalBranching}.

Both the FCFs and the HL factors may be calculated for CaOH using empirical methods. The transition dipole moment $|\mathbf{R}_e|^2$, meanwhile, is treated as constant for all rovibronic transitions in an electronic band, and may be factored out. This approximation is justified by the good agreement of calculated VBRs with experimental results (section \ref{sec:VBRfinal}), and by the practical value of calculations grounded solely in empirical parameters (determination of $|\mathbf{R}_e|^2$ would necessarily require some reference to \emph{ab initio} methods). Under this approximation, the branching ratio from an excited state $|\zeta',J'\rangle$ to a ground state $|\zeta'',J''\rangle$ is
\begin{equation}
    P_{|\zeta',J'\rangle \rightarrow |\zeta'',J''\rangle} \equiv \frac{\nu^3 q_{v'-v''} S^{J'}_{J''}(\zeta',\zeta'')}{\mathcal{N}} ,
\label{eqn:VBRcalc}
\end{equation}
where
\begin{equation}
\mathcal{N} = \sum_i \nu_i^3 q_{v'-v_i''} S^{J'}_{J''_i}(\zeta',\zeta''_i)
\end{equation}
is a normalization factor. The sum is over all ground states $|\zeta_i'',J_i''\rangle$ to which the initial state may decay.


In section \ref{sec:FCFcalc} below, we determine FCFs for CaOH by evaluating Eq. \ref{eqn:FCFintegral} in the harmonic approximation. In sections \ref{sec:RennerTeller} and \ref{sec:FermiResonance} we consider vibronic mixing due to Renner-Teller perturbations and Fermi resonance that significantly alters the results obtained from harmonic FCFs alone. In section \ref{sec:VBRfinal} we combine these effects to estimate VBRs for CaOH.

\begin{table}
\begin{tabular}{l c l | c c c}
\Xhline{1pt}
\\[-1 em]
\Xhline{1pt} 
\\[-0.7 em]
\multicolumn{1}{c}{Excited}& &\multicolumn{1}{c|}{Ground} & \multicolumn{1}{l}{$N''=1$} & \multicolumn{1}{l}{$N''=1$} & \multicolumn{1}{l}{$N''=2$} \\
\multicolumn{1}{c}{State} & & \multicolumn{1}{c|}{State} & $J''=1/2$ & $J''=3/2$ & $J''=3/2$ \\
\\[-1 em]
\Xhline{1pt}
\\[-.5 em]
$\widetilde{A}(000)^2\Pi_{1/2}$ & $\rightarrow$ & $\widetilde{X}(000)^2\Sigma^+$ & $2/3$ & $1/3$ & -- \\
$\widetilde{B}(000)^2\Sigma^+$ & $\rightarrow$ & $\widetilde{X}(000)^2\Sigma^+$ & $1/3$ & $2/3$ & -- \\
$\widetilde{A}(010)\mu^2\Sigma^{(+)}$ & $\rightarrow$ & $\widetilde{X}(01^10)^2\Pi$ & $0.108$ & $0.826$ & $0.066$ \\
$\widetilde{A}(010)\kappa^2\Sigma^{(+)}$ & $\rightarrow$ & $\widetilde{X}(01^10)^2\Pi$ & $0.559$ & $0.007$ & $0.434$ \\
$\widetilde{B}(01^10)^2\Pi$ & $\rightarrow$ & $\widetilde{X}(01^10)^2\Pi$ & $1/3$ & $1/6$ & $1/2$ \\
$\widetilde{A}(02^20)^2\Pi_{1/2}$ & $\rightarrow$ & $\widetilde{X}(02^20)^2\Delta$ & -- & -- & $1$ \\
\\[-1 em]
\Xhline{1pt} 
\\[-1 em]
\Xhline{1pt} 
\\[-0.5 em]
\end{tabular}
\caption{Degeneracy-weighted rotational line strengths, $S^{J'}_{J''}(\zeta',\zeta'')$, calculated for transitions from $J'=1/2, (+)$ parity excited states. The same factors hold for other vibrational levels of the $\widetilde{X}^2\Sigma^+$ ground state with similar vibronic symmetry. All ground states listed have $(-)$ parity. Calculations make use of dipole matrix elements found, e.g., in Ref. \cite{hirota2012high}.} 

\label{tab:RotationalBranching}
\end{table}

\subsection{FCF Calculations in the Harmonic Approximation}
\label{sec:FCFcalc}

Within the harmonic approximation, Eq. \ref{eqn:FCFintegral} may be evaluated analytically using empirically measured force constants, bond lengths, and vibrational frequencies for CaOH. We evaluate these integrals by first performing a classical Wilson GF-matrix analysis of the normal modes~\cite{wilson1955}, then evaluating the overlap integrals (including Duschinsky rotations) using the formalism of Sharp and Rosenstock~\cite{sharp1964franck, weber2003franck}. This method is the same as the one used in Ref. \cite{kozyryev2019determination} and is described in more detail in Appendix \ref{app:GFmatrix}. The resulting VBRs, accounting for the frequency and H\"onl-London scaling factors (Eq. \ref{eqn:VBRcalc}), are shown in the first column of Table \ref{tab:VBRComparison}. While this calculation accurately reproduces observed decays to the Ca--O stretching modes, it underestimates the observed $(02^00)$ bending mode decay by an order of magnitude and also fails to explain $\Delta \ell \neq 0$ transition strength. A more accurate description of these discrepancies requires consideration of additional perturbations.

\subsection{Renner-Teller Perturbation}
\label{sec:RennerTeller}
Vibrational branching to the $\widetilde{\text{X}}\left(01^10\right)$ and $\widetilde{\text{X}}\left(02^20\right)$ states is suppressed due to the nominal selection rule $\Delta \ell = 0$ due to the Born-Oppenheimer (BO) approximation. However, BO approximation breakdown at the level of a few parts per thousand is not unexpected. Transitions with $\lvert \Delta \ell \rvert = 1$ have been previously observed in SrOH~\cite{brazier1985laser, presunka1994laser}, BaOH~\cite{KinseyNielsen1986}, and CaOH~\cite{coxon1994laser, li1995bending}. In this work, we consider two decays in which $\lvert \Delta \ell \rvert \neq 0$: $\widetilde{\text{A}}^2\Pi_{1/2}\left(000\right) \rightarrow \widetilde{\text{X}}^2\Sigma^+\left(01^10\right)$ and $\widetilde{\text{A}}^2\Pi_{1/2}\left(000\right) \rightarrow \widetilde{\text{X}}^2\Sigma^+\left(02^20\right)$. These decays are due to Renner-Teller (RT) induced vibronic coupling. The Renner-Teller Hamiltonian has been covered in depth elsewhere~\cite{Aarts1978,  Brown1983, Jensen2000Renner, Brown2000RennerEffective, Brown2003, hirota2012high, brown1977effective} and is described in Appendix \ref{app:RTHam}.

There are three RT-induced vibronic coupling pathways relevant to this work: (1) direct RT mixing between $\widetilde{A}(000)$ and $\widetilde{B}(01^10)$; (2) second-order coupling between $\widetilde{A}(000)$ and $\widetilde{A}(01^10)$; and (3) direct RT mixing between $\widetilde{A}(000)$ and $\widetilde{A}(02^20)$. Each of these contributions is discussed in more detail below.

\subsection*{ \textbf{$\mathbf{ \widetilde{\text{A}}^2}\Pi\mathbf{_{1/2}\left(000\right) \rightarrow \widetilde{\text{X}}^2}\Sigma\mathbf{^+\left(01^10\right)}$ decay}}

The $\lvert \Delta \ell \rvert=1$ transition has been previously observed in several alkaline-earth monohydroxides~\cite{brazier1985laser, presunka1994laser, coxon1994laser, li1995bending, KinseyNielsen1986}. This transition could gain strength due to direct vibronic coupling between $\widetilde{\text{B}}^2\Sigma^+\left(01^10\right)$ and $ \widetilde{\text{A}}^2\Pi_{1/2}\left(000\right)$, or through a second-order spin-orbit vibronic coupling~\cite{li1997cacch}. Previous authors have favored the latter explanation based on the observation that the intensity grows with atomic spin-orbit constant~\cite{brazier1985laser, coxon1994laser, presunka1994laser}. In this work we find that both contributions are necessary to satisfactorily explain the observed $\lvert \Delta \ell \rvert=1$ branching in CaOH.


The direct vibronic mixing between $\widetilde{A}(000)$ and $\widetilde{B}(01^10)$ may be evaluated in perturbation theory as
\begin{align}
\lvert \langle \widetilde{B}(\overline{01^10})^2\Pi \lvert \widetilde{A}(000)^2\Pi_{1/2} \rangle \rvert^2 &\approx \frac{\lvert \langle \widetilde{B}\lvert H_\text{RT} | \widetilde{A} \rangle \rvert ^2}{\left(\Delta E_{AB} - \omega_2\right)^2} \nonumber \\
&\approx \frac{g_K}{\omega_2\left(1-\frac{\omega_2}{\Delta E_{AB}}\right)^2} \nonumber \\
&\approx 1.2 \times 10^{-3}
\label{eqn:dipolarRT010}
\end{align}
where $\Delta E_{AB} = E(\widetilde{A}) - E(\widetilde{B})$ is the splitting between the $\widetilde{A}$ and $\widetilde{B}$ state energy origins, $\omega_2$ is the bending vibrational frequency, and the strength of the relevant RT interaction is contained in the spectroscopic parameter $g_K$ (see Appendix \ref{app:RTHam}). A bar is used to specify unperturbed basis states, which are eigenstates of the molecular Hamiltonian absent the perturbations considered here. All spectroscopic quantities were measured in Refs. \cite{li1995bending, li1996dye, bernath1984dye} and are compiled in Table \ref{tab:Constants}.

\begin{table}
\begin{tabular}{l l c}
\Xhline{1pt}
\\[-1 em]
\Xhline{1pt} 
\\[-0.7 em]
Parameter & Value (cm$^{-1}$) & Ref. \\
\\[-1 em]
\Xhline{1pt}
\\[-.5 em]
$A_{SO}$ & 66.8181 & \cite{li1995bending} \\
$g_K$ & 0.5937 & \cite{li1995bending} \\
$\omega_2$ & 366.435 & \cite{li1996dye} \\
$\epsilon \omega_2$ & -36.5641 & \cite{li1996dye} \\
$\Delta E_{AB}$ & -2024.14 & \cite{li1995bending,bernath1984dye} \\
$\Delta E_{(020)}$ & -702.051 & \cite{li1995bending,li1996dye} \\
$g_{22}$ & 7.5314 & \cite{li1996dye} \\
\\[-1 em]
\Xhline{1pt} 
\\[-1 em]
\Xhline{1pt} 
\\[-0.5 em]
\end{tabular}
\caption{Compiled spectroscopic parameters, as defined in the text, used in the branching ratio calculations in this work.} 

\label{tab:Constants}
\end{table}

Vibronic mixing between $\widetilde{A}(000)$ and $\widetilde{A}(01^10)$ is induced by the $H_\text{RT} \times H_\text{SO}$ interaction, where $H_\text{SO}$ is the spin-orbit Hamiltonian. This second-order interaction proceeds via both the $\widetilde{B}(000)^2\Sigma^+$ and the $\widetilde{B}(01^10)^2\Pi$ states, and the fractional mixing is given by (see Appendix \ref{app:RTHam})
\begin{align}
    \lvert \langle \widetilde{A}(\overline{010})^2\Sigma\lvert \widetilde{A}(000)^2\Pi_{1/2} \rangle \rvert ^2 &\approx 2 \left | \frac{ 2 \langle \widetilde{A} \lvert H_\text{SO} \rvert \widetilde{B}\rangle \langle \widetilde{B} \lvert H_\text{RT} \rvert \widetilde{A}\rangle}{\omega_2 \Delta E_{AB}} \right | ^2 \nonumber \\
    &\approx \frac{4 g_K A_\text{SO}^2}{\omega_2^3} \approx 2.2 \times 10^{-4}
\label{eqn:RTSO010}
\end{align}
where $A_\text{SO}$ is the spin-orbit constant in the $\widetilde{A}^2\Pi$ state. The factor of 2 in the numerator of the first line accounts for contributions via the $\widetilde{B}(000)$ and $\widetilde{B}(01^10)$ states, which constructively interfere. An additional, overall factor of 2 arises due to the presence of two distinct vibronic components of the $\widetilde{A}^2\Pi(010)$ vibrational manifold (see App. \ref{app:structure} and App. \ref{app:RTHam}). These components -- labeled $\mu^2\Sigma^{(+)}$ and $\kappa^2\Sigma^{(-)}$ \cite{li1995bending}, where $\mu$ and $\kappa$ denote the lower- and higher-energy states, respectively -- are split via the RT and spin-orbit interactions and both mix with $\widetilde{A}(000)^2\Pi$, approximately doubling the total $\widetilde{A}(000)-\widetilde{A}(01^10)$ admixture. Though this second-order $\widetilde{A}(000)-\widetilde{A}(01^10)$ mixing is still a factor of $\sim 5$ weaker than direct $\widetilde{A}(000)-\widetilde{B}(01^10)$ mixing in CaOH, it is expected to dominate for the heavier alkaline earth monohydroxides, which have weaker vibronic coupling but a significantly stronger spin-orbit interaction \cite{presunka1994laser, KinseyNielsen1986}.

Applying the scaling factors of Eq. \ref{eqn:VBRcalc}, for the $N''=1$ rotational component of $\widetilde{\text{X}}^2\Sigma^+(01^10)$ we arrive at a branching ratio of
\begin{align}
    P_{(000)-(01^10)} &\approx \frac{\nu^3}{\mathcal{N}} \nonumber \\
    \times \sum_{\substack{N''=1 \\ J''=1/2,3/2}}&\bigg|\sum_{\zeta'} \langle \widetilde{A}(000)^2\Pi_{1/2} \lvert \zeta' \rangle \langle \zeta' \lvert \widetilde{X}(010) \rangle \langle \mu \rangle^{\zeta'}_{N'',J''} \bigg|^2 \nonumber \\
    &\approx 3.4\times 10^{-4}
\label{eqn:010VBR}
\end{align}
where the second sum is over the excited vibronic states $\zeta'=$ $\widetilde{A}(\overline{010})\mu$, $\widetilde{A}(\overline{010})\kappa$, and $\widetilde{B}(\overline{010})$ (other states with $\ell\neq 1$ are symmetry-forbidden). Each term is the product of a vibronic mixing amplitude $\langle \widetilde{A}(000)^2\Pi_{1/2} \lvert \zeta' \rangle$, a harmonic FCF $\langle \zeta' \lvert \widetilde{X}(\overline{010}) \rangle$, and a rotational factor $\langle \mu \rangle ^{\zeta'}_{N'',J''}$. The relative rotational transition intensities are given by the H\"onl-London factors, $S^{J'}_{J''}(\zeta',\zeta'') \equiv |\langle \mu \rangle ^{\zeta'}_{N'',J''}|^2$, shown in Table \ref{tab:RotationalBranching}. The VBR is less than the sum of the mixing probabilities in Eqs. \ref{eqn:dipolarRT010} and \ref{eqn:RTSO010} because only decay to $N''=1$ is considered ($N''=2$ contributes another $\sim 3 \times 10^{-4}$), and because interference between terms in the summation of Eq. \ref{eqn:010VBR} reduces the total.

Performing similar calculations for the $\widetilde{B}^2\Sigma^+(000)$ state of CaOH, we determine VBRs (dominated by direct RT mixing with the $\widetilde{A}(010)^2\Sigma$ states) of $2.5 \times 10^{-3}$ to $N''=1$ and $3.6 \times 10^{-4}$ to $N''=2$. This agrees very well with the measured (rotationally-unresolved) VBR of $3 \times 10^{-3}$ from previous dispersed-fluorescence experiments \cite{kozyryev2019determination}. Because $\sim$$4.3\%$ of the total photons scattered in the experimental cycling scheme (Sec. \ref{VBRmeasurements}) are from $\widetilde{B}(000)$, we therefore calculate an ``effective'' VBR for $\widetilde{X}(01^10), N''=1$ decay of $0.957\times (3.4\times 10^{-4}) + 0.043\times(2.5 \times 10^{-3}) \approx 4.3 \times 10^{-4}$.

\subsection*{\textbf{$\mathbf{ \widetilde{\text{A}}^2}\Pi\mathbf{_{1/2}\left(000\right) \rightarrow \widetilde{\text{X}}^2}\Sigma\mathbf{^+\left(02^20\right)}$ decay}}

The $\lvert \Delta \ell \rvert = 2$ transition gains transition strength due to mixing of the $\widetilde{\text{A}}^2\Pi_{1/2}\left(000\right)$ state with the $\widetilde{\text{A}}^2\Pi_{1/2}\left(02^20\right)$ state by the Renner-Teller Hamiltonian. The quadrupolar term in $H_\text{RT}$ can couple states according to the selection rules $\Delta \Lambda = \pm 2$, $\Delta \ell = \mp 2$, $\Delta \Sigma = 0$. This term therefore mixes certain components of the $\widetilde{\text{A}} \, ^2\Pi(000)$ and $\widetilde{\text{A}} \, ^2\Pi(020)$ states at first order. (The dipolar term of $H_\text{RT}$ also contributes at second order to this mixing \cite{brown1977effective}.) A very similar interaction has previously been observed between $\widetilde{\text{A}} \, ^2\Pi(01^10)$ and $\widetilde{\text{A}} \, ^2\Pi(03^30)$ in CaOH~\cite{li1995bending}.

The relevant interactions are parametrized by the quantity $\epsilon\omega_2$, where $\epsilon$ is the Renner parameter. The matrix element connecting the $\widetilde{A}(000)$ and $\widetilde{A}(02^20)$ states is $\epsilon \omega_2/\sqrt{2}$ \cite{li1995bending}, and the resulting admixture is
\begin{align}
  \lvert \langle \widetilde{\text{A}}(\overline{02^20})&^2\Pi_{1/2} \lvert \widetilde{\text{A}}(000) ^2\Pi_{1/2} \rangle \rvert^2 \nonumber \\
  &\approx \frac{(\epsilon \omega_2/\sqrt{2})^2}{(\Delta E_{(020)}-4g_{22})^2} \approx 1.2 \times 10^{-3}
\end{align}
where $\Delta E_{(020)} = E_{(000)}- E_{(020)}$ is the  difference between the $\widetilde{A}(000)$ and $\widetilde{A}(02^00)$ origin energies, and $g_{22}\ell^2$ is the energy shift of a state with vibrational angular momentum $\ell\neq 0$.

Applying Eq. \ref{eqn:VBRcalc}, the VBR for decay from $\widetilde{A}(000)^2\Pi_{1/2} \rightarrow \widetilde{X}(02^20)^2\Delta$ is therefore
\begin{align}
    P_{(000)-(02^20)} \approx &\frac{\nu^3}{\mathcal{N}}\lvert \langle \widetilde{\text{A}}(\overline{02^20})^2\Pi_{1/2} \lvert \widetilde{\text{A}}(000) ^2\Pi_{1/2} \rangle \rvert^2 \nonumber \\
    &\times S^{1/2}_{3/2}(\widetilde{A}(02^20), \widetilde{X}(02^20)) q_{(02^20)-(02^20)} \nonumber \\ 
    \approx &1.0 \times 10^{-3}
\end{align}
where $q_{(02^20)-(02^20)}\approx 0.95$ is the harmonic FCF defined in Eq. \ref{eqn:FCFintegral}, and all decay is to the $(N''=2, J''=3/2, -)$ rotational level.

\subsection{Fermi Resonance}
\label{sec:FermiResonance}

Because the Ca--O stretching frequency is approximately double that of the Ca--O--H bending frequency in CaOH~\cite{li1996dye}, vibrational states differing by $\Delta v_1 = \pm 1$, $\Delta v_2 = \mp 2$ come in closely spaced groups, e.g., $(100)\sim(020)$ and $(200)\sim(120)\sim(040)$ (see Fig. \ref{fig:Xvibstructure} in Appendix \ref{app:structure}). States of the same vibronic symmetry within these polyads are mixed by a cubic term in the potential energy surface of the form $V_F = k_{122}Q_1Q_2^2$ \cite{fermi1931ramaneffekt, hougen1962fermi}, where $Q_1$ and $Q_2$ are normal coordinates associated with Ca--O stretching and Ca--O--H bending, respectively. This leads to intensity borrowing which can significantly affect the magnitude of vibronic decay channels including $\widetilde{\text{X}}\,^2\Sigma^+(02^00)$ and $\widetilde{\text{X}}\,^2\Sigma^+(12^00)$. This effect is well understood, and is known as the Fermi resonance interaction~\cite{fermi1931ramaneffekt, hougen1962fermi}.

The Fermi resonance interaction is quantified by a parameter $W$, which is proportional to $k_{122}$ and parametrizes the strength of the cubic mixing (see Appendix \ref{app:Fermi}). It is estimated that $W \approx -10.7$ cm$^{-1}$ in the $\widetilde{\text{X}}\,^2\Sigma^+$ state based on prior measurements in CaOH~(see Appendix \ref{app:Fermi}). Including this off-diagonal coupling within the $\widetilde{\text{X}}(100)\sim(02^00)$ Fermi dyad leads to the following mixing amplitudes:
\begin{align}
    |\widetilde{\text{X}}(100)\rangle &= 0.96|\widetilde{\text{X}}(\overline{100})\rangle + 0.28|\widetilde{\text{X}}(\overline{02^00})\rangle \nonumber \\
    |\widetilde{\text{X}}(02^00)\rangle &= -0.28|\widetilde{\text{X}}(\overline{100})\rangle + 0.96|\widetilde{\text{X}}(\overline{02^00})\rangle.
\label{eqn:Fermimixingamp}
\end{align}
where a bar is used to denote the unperturbed harmonic oscillator basis states. The VBR from $\widetilde{\text{A}}^2\Pi_{1/2}(000) \rightarrow \widetilde{\text{X}}^2\Sigma^+(02^00)$ is then given by
\begin{align}
P_{(000)-(02^00)}& \nonumber \\
\approx \big| 0.96 &\langle \widetilde{\text{X}}(\overline{02^00})  | \widetilde{\text{A}}(\overline{000})\big \rangle - 0.28 \langle \widetilde{\text{X}}(\overline{100}) |  \widetilde{\text{A}}(\overline{000})\big \rangle \big|^2 \nonumber \\
&\times \frac{\nu^3 (S_{3/2}^{1/2}+S_{1/2}^{1/2})}{\mathcal{N}} \approx 4.6 \times 10^{-3},
\end{align}
where each inner product on the right-hand side is a harmonic overlap integral that can be evaluated using the methods of section \ref{sec:FCFcalc}, and the H\"onl-London factors are those for $\ell = 0$ states in Table \ref{tab:RotationalBranching}. A similar expression applies to the $\widetilde{\text{X}}^2\Sigma^+(100)$ branching ratio. The Fermi resonance mixing therefore increases the branching ratio to $\widetilde{\text{X}}(02^00)$ by an order of magnitude compared to the purely harmonic result.

Fermi mixing between $\widetilde{\text{X}}(200)$ and $\widetilde{\text{X}}(12^00)$ is also expected to cause significant branching to the $\widetilde{\text{X}}\,^2\Sigma^+(12^00)$ state. This is predicted to be one of the primary loss channels from the photon cycling scheme in Fig. \ref{fig:cyclingscheme}.

\subsection{Calculation Results}
\label{sec:VBRfinal}

Table \ref{tab:VBRComparison} compares calculated branching ratios with the experimental results of section \ref{VBRmeasurements}. The calculations generally exhibit good agreement with measured branching ratios. While branching to the Ca--O stretching modes $(000)$, $(100)$, and $(200)$ is dominated by harmonic wavefunction overlap, decay to the bending modes arises predominantly from other mechanisms. Branching to $\widetilde{\text{X}}\,^2\Sigma^+(02^00)$ occurs primarily due to Fermi resonance mixing with the $\widetilde{\text{X}}\,^2\Sigma^+(100)$ state, while decay to the $\ell\neq 0$ bending modes $(02^20)$ and $(01^10)$ arises from vibronic perturbations in the $\widetilde{\text{A}}$ and $\widetilde{\text{B}}$ states, as described in section \ref{sec:RennerTeller}. The largest remaining disagreement is in decay to the $(200)$ stretching mode, which is likely enhanced due to anharmonic terms in the potential energy surface not included in these calculations \cite{koput2002AbInitio}. All other values agree within a factor of 2 with their experimental counterparts and may be regarded as useful predictors of the relative significance of vibrational branching pathways.

For most states the $\widetilde{\text{A}}\rightarrow \widetilde{\text{X}}$ VBR is a good approximation to the observed branching out of the experimental cycling scheme, which employs both $\widetilde{\text{X}}\,^2\Sigma^+(000)\rightarrow\widetilde{\text{A}}\,^2\Pi_{1/2}(000)$ and $\widetilde{\text{X}}\,^2\Sigma^+(100)\rightarrow\widetilde{\text{B}}\,^2\Sigma^+(000)$ transitions. One notable exception, however, is the $\widetilde{\text{X}}(01^10)$ state, which experiences significant decay from $\widetilde{\text{B}}(000)$ with a VBR of $3(1) \times 10^{-3}$~\cite{kozyryev2019determination}. To account for this, we scale the separately-calculated $\widetilde{\text{A}}\rightarrow \widetilde{\text{X}}$ and $\widetilde{\text{B}}\rightarrow \widetilde{\text{X}}$ VBRs by the relative number of photon scattering events through each excited state \cite{kozyryev2019determination}. The effective branching ratio out of the experimental cycling scheme is therefore approximately
\begin{equation}
    P_\text{eff} \approx 0.957P_{\widetilde{\text{A}}\rightarrow \widetilde{\text{X}}} + 0.043P_{\widetilde{\text{B}}\rightarrow \widetilde{\text{X}}}
\end{equation}
since on average $\sim$95.7\% ($\sim$4.3\%) of the surviving molecular population is excited from $\widetilde{\text{X}}(000)$ ($\widetilde{\text{X}}(100)$) based on the measured branching ratios. This correction is significant only for the $\widetilde{\text{X}}(01^10)$ state, though makes small contributions to the other VBRs included in Table \ref{tab:VBRComparison}.

The final row of Table \ref{tab:VBRComparison} compares the calculated and measured total branching ratio to states not addressed by the photon cycling scheme of Fig. \ref{fig:cyclingscheme}. We may use the calculations to make qualitative predictions about the states that must be addressed to cycle up to $\sim$$10^4$ photons in CaOH. The calculated loss is dominated by decay to the unaddressed $N''=2$ component of $\widetilde{\text{X}}(01^10)$ at the $\sim 3$$\times10^{-4}$ level; and by loss to $\widetilde{\text{X}}\,^2\Sigma^+(12^00)$ at $\sim$$1.5\times10^{-4}$ due to Fermi resonance mixing with $\widetilde{\text{X}}(200)$. In anticipation of the need to repump the latter state, high resolution spectroscopy of the (120) manifold was performed (see Appendix \ref{spectroscopy}).

Additional loss channels may be more significant than these calculations indicate, gaining strength through perturbations neglected in the analysis. Specifically, though decay to $\widetilde{\text{X}}(300)$ is predicted to occur at the $\sim$$10^{-6}$ level based on harmonic calculations, it is likely that anharmonicity in the potential energy surface will significantly enhance loss to this state. Preliminary experimental measurements indeed suggest that the VBR to $\widetilde{\text{X}}(300)$ may be as large as $\sim$$1\times 10^{-4}$. Finally, decay to the O--H stretching mode may be significant at this level, as experimental uncertainty in the O--H bond length makes calculations for this state unreliable.

\begin{figure}
    \centering
    \includegraphics[width = \linewidth]{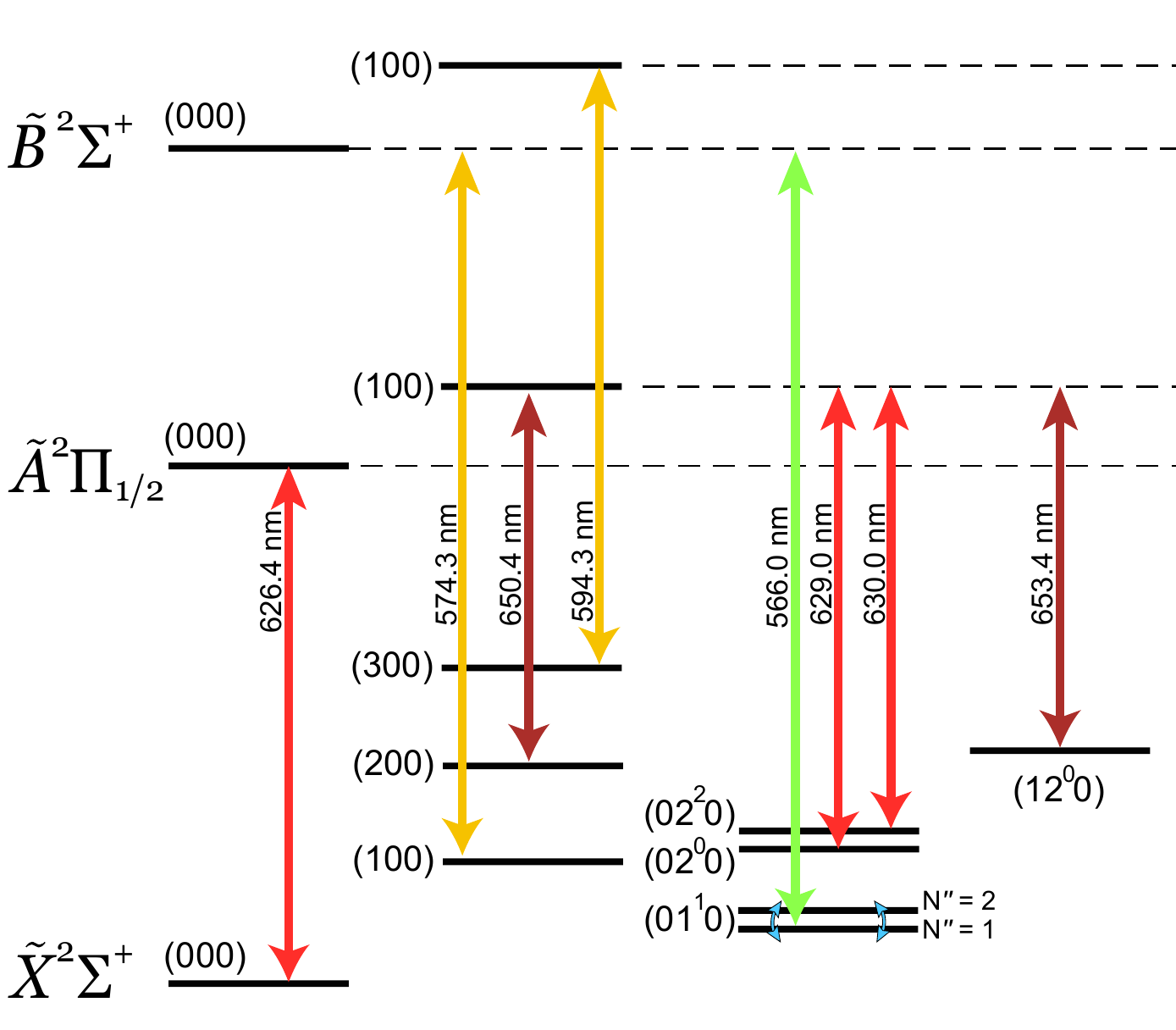}
    \caption{Photon cycling scheme and vibrational structure of CaOH to achieve scattering of $\sim$ 5000 photons. All arrows represent transitions coupled by laser light. The $N''=1$ and $N''=2$ rotational components of the $\widetilde{\text{X}}\,^2\Sigma^+(01^10)$ state may either be coupled by 40~GHz microwave radiation or directly addressed by frequency modulation of the laser radiation.}
    \label{fig:finalscheme}
\end{figure}

Taken together, these calculations predict that 2--3 additional repumping lasers, in addition to remixing of the $\widetilde{\text{X}}(01^10)(N''=1,2)$ levels via microwave radiation or laser frequency modulation, will be necessary to scatter $\sim$$5000$ photons in CaOH before significant population loss to dark states. One proposed cycling scheme is illustrated in Fig. \ref{fig:finalscheme}. The calculations presented above predict that it should be possible to scatter 4700 photons per molecule on average with this combination of repumping lasers. This should be sufficient to enable radiative slowing and magneto-optical trapping of CaOH molecules if loss of some molecules to dark states is tolerated.

\vspace{-0.25 cm}

\section{Conclusion}

This work establishes a vibrational repumping scheme enabling deep laser cooling and control of CaOH despite its complex internal structure. Our experimental measurements validate calculations used to estimate higher-order decay pathways. Using these experimentally-validated predictions, we propose a laser cooling scheme (shown in Fig. \ref{fig:finalscheme}) capable of scattering on the order of $\sim$5000 photons per molecule on average, which should enable experimental efforts to implement radiative slowing, 3D magneto-optical trapping, and ultimately, deep laser cooling into the ultracold regime. The spectroscopy presented in Appendix \ref{spectroscopy} establishes a clear path towards the experimental implementation of this laser cooling scheme. While the measurements presented here are unique to CaOH, the experimental methods and calculations outline a general framework that could be used to predict and confirm branching ratios in other molecular candidates for direct laser cooling. The insights gained here for CaOH can be generalized to support recent proposals extending laser cooling to symmetric and asymmetric top molecules~\cite{mitra2020direct,augenbraun2020molecular,kozyryev2016proposal,dickerson2020Franck,augenbraun2020observation} and even molecules with multiple cycling centers~\cite{o2019hypermetallic,ostojic2013predicted,Ivanov2020,klos2020prospects}.

\begin{acknowledgements}
We would like to thank L. Anderegg and Z. Lasner for insightful discussions. This work was supported by the NSF, AFOSR, and ARO. N.B.V. acknowledges funding from the NDSEG fellowship, and B.L.A. from the NSF GRFP.
\end{acknowledgements}

\appendix

\section{Experimental Sequence for Section \ref{sec:HighVBR}}
\label{app:ExpSeq}

The experimental sequence for these measurements interleaves experimental conditions to normalize against fluctuations in molecular number that are common for ablation-based production. This sequence also determines the non-negligible natural population present in the excited vibrational modes and the results of imperfect optical pumping. Collecting data in four different experimental conditions allowed us to measure and correct for these factors and extract the percentage of recovered population ($P_{rec}$). These conditions are indicated below, where ``interaction'', ``cleanup'', and ``detection'' refer to the regions of optical access depicted in \autoref{fig:experimentaldiagram}(b).

\begin{itemize}
    \item[] \textbf{A} \qquad Detection Light only
    \item[] \textbf{B} \qquad Detection Light and Cleanup Light
    \item[] \textbf{C} \qquad Interaction Light and Detection Light
    \item[] \textbf{D} \qquad Interaction, Cleanup, and Detection light
\end{itemize}

These four measurement configurations allow us to extract the population optically pumped into higher vibrational states (\textbf{A}-\textbf{C}), the natural population in higher vibrational states (\textbf{B}-\textbf{A}), and the population recovered from excited vibrational states (\textbf{D}-\textbf{C}-(\textbf{B}-\textbf{A})), all while normalizing against ablation fluctuations. The recovered population, $P_{rec}$, is the percentage of depleted population that is recovered upon repumping one (or many) excited vibrational states and calculated as:

\begin{equation}
   P_{rec} = \frac{\text{\textbf{D} - \textbf{C} - (\textbf{B}-\textbf{A})}}{\text{\textbf{A} - \textbf{C}}}.
\end{equation}

\begin{figure}[h]
    \centering
    \includegraphics[width = \linewidth]{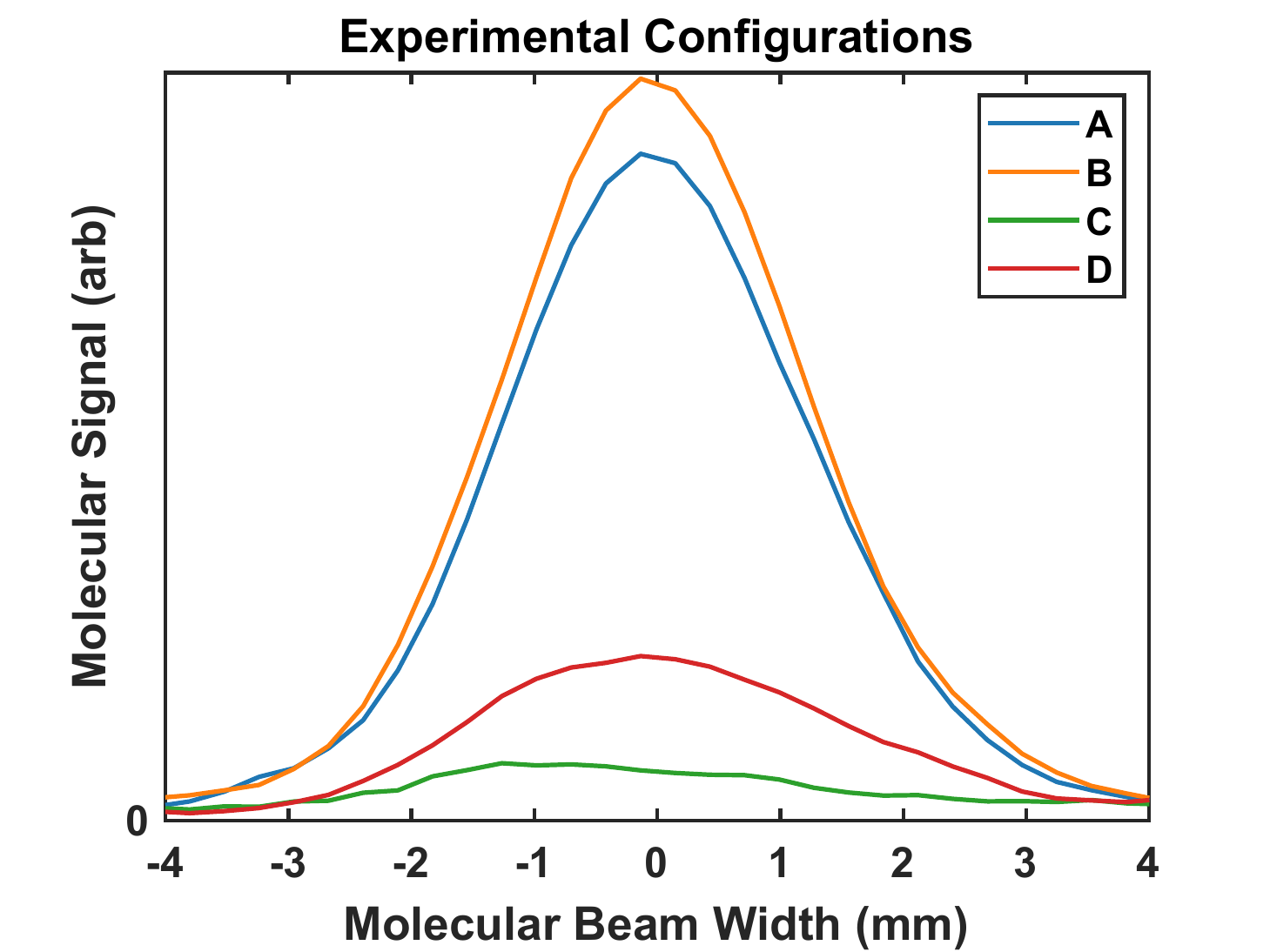}
    \caption{Molecular Signal under the four experimental configurations to determine the repump fraction of $\widetilde{\text{X}}(01^10)$. This depicts how substantial fractions of molecular population may be optically pumped via small decay pathways during repeated photon cycling.}
    \label{fig:allExpConditions}
\end{figure}

An example of the data collected from this experimental sequence is given in Fig. \ref{fig:allExpConditions}, where each of the experimental conditions is indicated.

\section{Supplementary Level Diagrams}
\label{app:structure}

\begin{figure}
    \centering
    \includegraphics[width = 0.9\columnwidth]{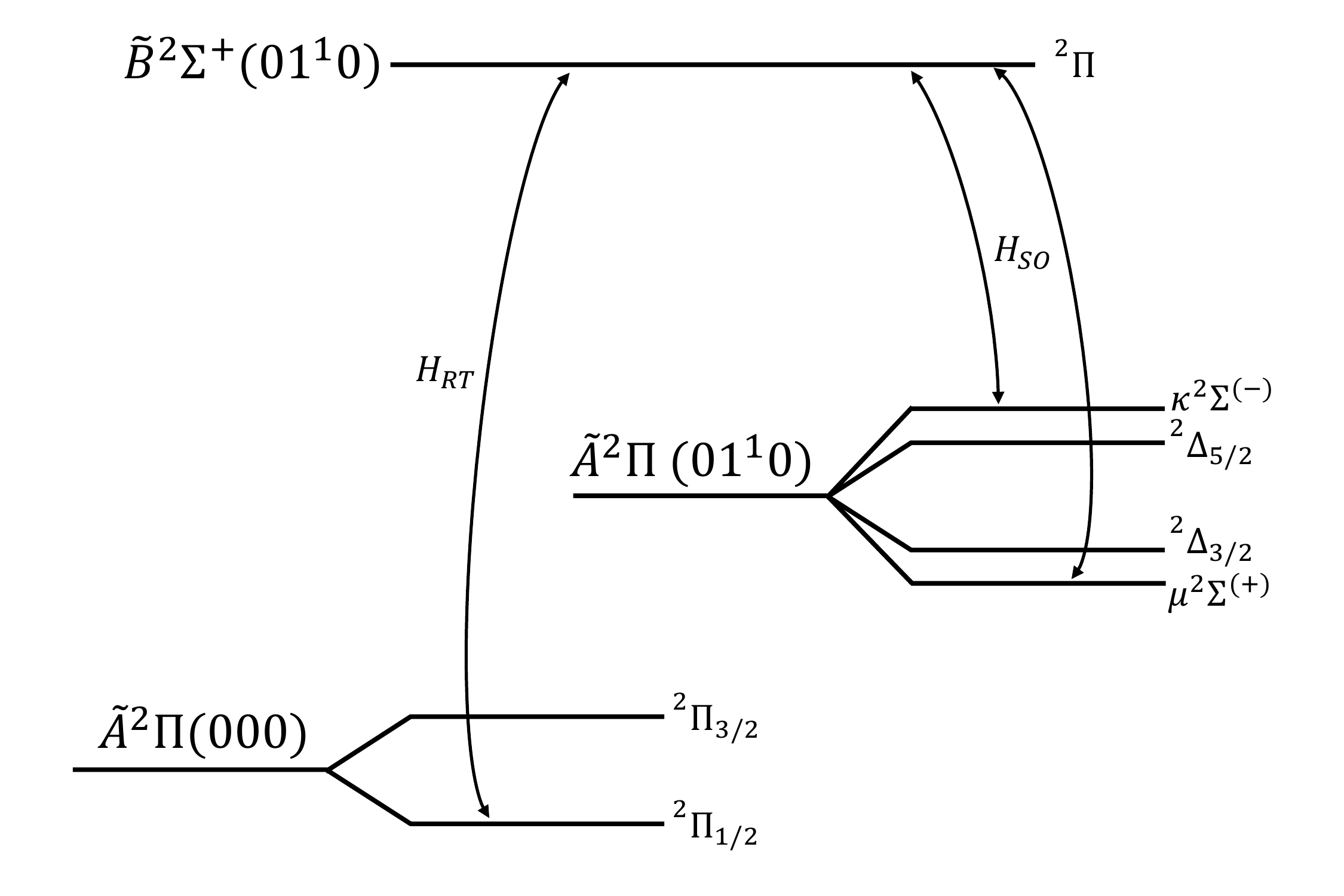}
    \caption{Illustration of the processes leading to vibronic mixing of $\widetilde{A}^2\Pi(000)$ with states of $\ell =1$ character. First-order mixing due to the dipolar term of $H_\text{RT}$ directly connects $\widetilde{A}(000)^2\Pi_{1/2} - \widetilde{B}(01^10)^2\Pi$. Additionally, second-order mixing between the $\widetilde{A}(000)$ and $\widetilde{A}(01^10)$ vibrational manifolds occurs when the spin-orbit interaction, $H_\text{SO}$, is also considered. This second order process occurs via the $\widetilde{B}(01^10)^2\Pi$ state, as well as through $\widetilde{B}^2\Sigma^+(000)$ via an analogous process (not pictured). Admixtures of both the $\mu^2\Sigma^{(+)}$ and $\kappa^2\Sigma^{(-)}$ \cite{hougen1962rotational,li1995bending} components of the $\widetilde{A}(010)$ vibronic manifold appear due to this process. Energies are not shown to scale.}
    \label{fig:A010mixing}
\end{figure}

\begin{figure}
    \centering
    \includegraphics[width = 0.9\columnwidth]{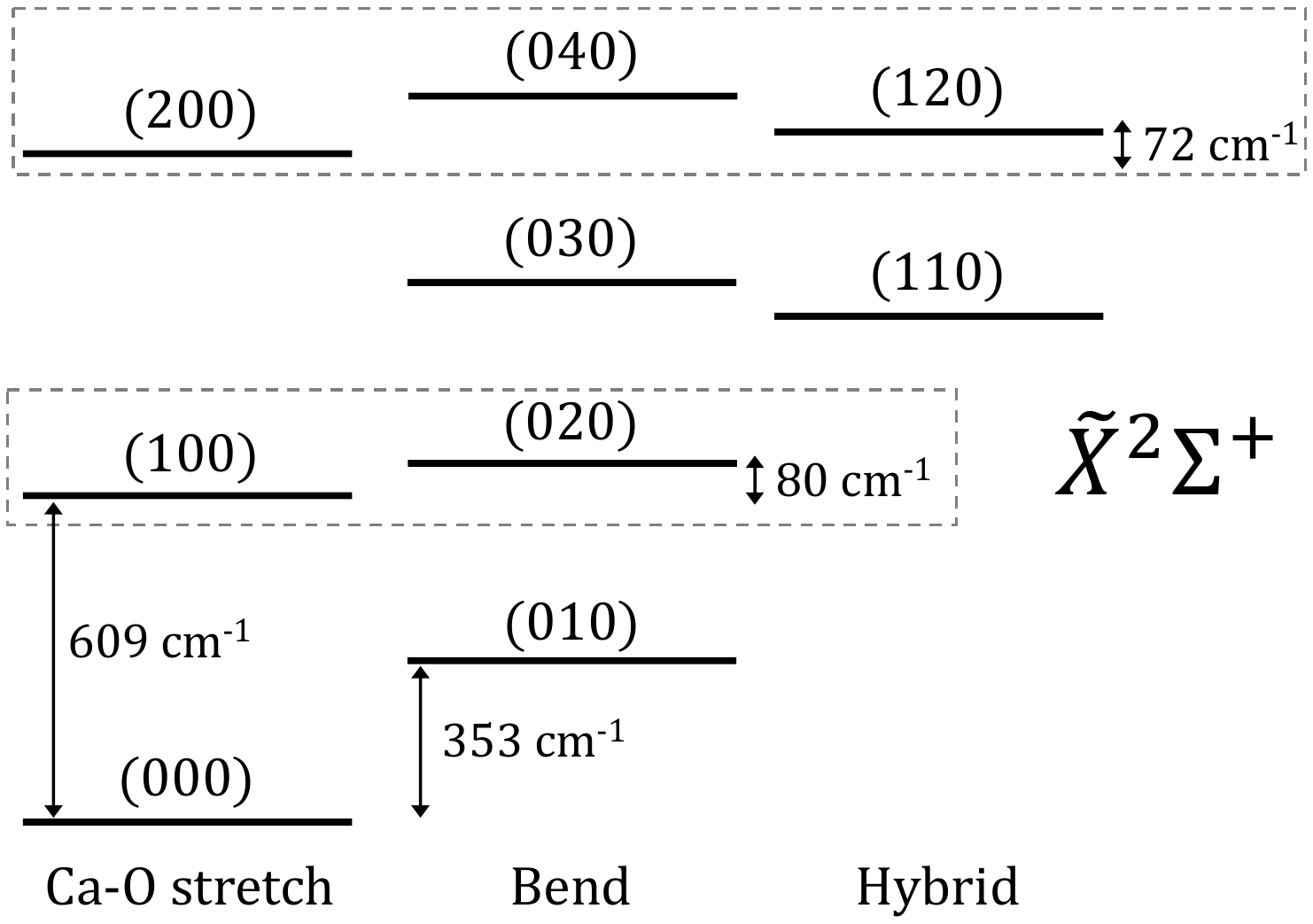}
    \caption{Low-lying vibrational levels of the $\widetilde{X}^2\Sigma^+$ state of CaOH. Groups of near-degenerate states with significant Fermi resonance mixing considered in this work are indicated by gray boxes. Note that Fermi resonance between $(11^10)$ and $(03^10)$ may also occur but is not considered here.}
    \label{fig:Xvibstructure}
\end{figure}

Level diagrams illustrating the Renner-Teller and Fermi resonance mixings discussed in the text are provided in Figs. \ref{fig:A010mixing}, and \ref{fig:Xvibstructure}.

\section{Spectroscopy of Combination Bands}
\label{spectroscopy}

While high-resolution spectroscopy exists for the $(300)$, $(400)$, and $(01^10)$ vibrational manifolds of the $\widetilde{\text{X}}\,^2\Sigma^+$ state~\cite{coxon1992investigation, li1995bending}, there is no such data for the $(120)$ manifold required to achieve a MOT of CaOH.\footnote{This manifold was, however, observed at low resolution in Ref. \cite{pereira1996observation}.} Therefore, high-resolution spectroscopy of the $\widetilde{\text{X}}(12^00)$ and $\widetilde{\text{X}}(12^20)$ repumping transitions was performed here.

Population of the $\widetilde{\text{X}}(12^00)$ ($\widetilde{\text{X}}(12^20)$) state was enhanced via off-diagonal vibronic decay after exciting molecules on the $\widetilde{\text{X}}(000)\rightarrow \widetilde{\text{A}}(02^00)$ ($\widetilde{\text{X}}(000)\rightarrow \widetilde{\text{A}}(02^20)$)  $P_{11}$ ($J''=3/2$) and $^PQ_{12}$ ($J''=1/2$) transitions.\footnote{For notational clarity, we label the $\widetilde{\text{A}}$ state vibrational levels by their dominant character. Specifically, the correspondence is $\widetilde{\text{A}}(02^00) \rightarrow \widetilde{\text{A}}(020)\mu^2\Pi_{1/2}$ and $\widetilde{\text{A}}(02^20) \rightarrow \widetilde{\text{A}}(020)\kappa^2\Pi_{1/2}$ in the notation of Ref. \cite{li1996dye}.} This ensured that only the $(N''=1, J''=1/2,3/2)$ rotational states relevant to laser cooling would be populated. A cw dye laser was then scanned over the repumping transition of interest. Several rovibronic transitions were observed; the absolute frequency of these transitions is reported in \autoref{tab:frequencies}. The relative frequency of the excitation laser was referenced to a High Finesse WS7 wavemeter. Empirically, the relative accuracy of the wavemeter has been verified to < 1 MHz when continuously calibrated with an atomic reference~\cite{loicthesis}. The absolute frequency was obtained by referencing the observed calcium $4s^2\,^1S_0 \rightarrow 4s4p\,^3P_1$ intercombination line to the accepted literature value~\cite{Degenhardt2005calcium}. The frequency offset (relative frequency -- absolute frequency) is assumed to be constant, and this constant correction is applied to the transitions reported in \autoref{tab:frequencies}. Several other known atomic and molecular transitions have been observed using this frequency reference, and the standard deviation of the frequency offsets from these measurements is $\sim$ 150 MHz. Rotational assignments were verified by selectively populating only specific rotational ground states as described above, and by confirming that the observed rotational spacings and spin-orbit splittings were consistent with the assignments.

\begin{table}
\begin{tabular}{ccccc}
\Xhline{1pt}
\\[-1 em]
\Xhline{1pt}
\\[-1 em]
&\multicolumn{4}{c}{$\widetilde{\text{A}}(100)$ $\leftarrow$ $\widetilde{\text{X}}(12^00)$}\\
$J$ & $P_{11}$ & $^PQ_{12}$ &  $Q_{11}$ & $^QR_{12}$ \\
\Xhline{1pt}
1/2 & & 458.789969 && 458.818297\\
3/2 & 458.789916 && 458.818244&\\
&&&&\\[-0.5 em]
&\multicolumn{4}{c}{$\widetilde{\text{A}}(02^00)$ $\leftarrow$ $\widetilde{\text{X}}(12^00)$}\\
\\[-1 em]
$J$ & $P_{11}$ & $^PQ_{12}$ &  & \\
\Xhline{1pt}
1/2 & &460.679181&&\\
3/2 & 460.679123&&&\\
&&&&\\[-0.5 em]
&\multicolumn{4}{c}{$\widetilde{\text{A}}(100)$ $\leftarrow$ $\widetilde{\text{X}}(12^20)$}\\
$J$ & $^OP_{12}$ &  &  $^PQ_{12}$ &\\
\Xhline{1pt}
3/2 & 458.180549 && 458.208878 &\\
&&&&\\[-0.5 em]
&\multicolumn{4}{c}{$\widetilde{\text{A}}(02^00)$ $\leftarrow$ $\widetilde{\text{X}}(12^20)$}\\
$J$ & $^OP_{12}$ &&& \\
\Xhline{1pt}

3/2 & 460.069796 &&&\\
[0.5 em]
\Xhline{1pt}
\\[-1 em]
\Xhline{1pt}
\\[-0.5 em]
\end{tabular}
\caption{Transitions observed during this work, reported in units of THz. The uncertainty in absolute frequency ($\sim$150 MHz) is estimated by comparing several measured transitions to accepted literature values, and applying a constant frequency correction to compensate. The remaining error is due to variation in these offsets as a function of wavelength, and is expected to manifest itself as an additional uniform shift of all frequencies reported here. The error in relative frequency ($\sim$10 MHz) is limited by the fit to the Doppler broadened lineshape.}
\label{tab:frequencies}
\end{table}

\section{FCF Calculations}
\label{app:GFmatrix}

The vibrational overlap integrals of Eq. \ref{eqn:FCFintegral} are calculated by assuming harmonic oscillator wavefunctions. The calculation employs the Wilson GF normal coordinate analysis and the Sharp-Rosenstock method, as described in detail elsewhere~\cite{wilson1955, sharp1964franck, weber2003franck, kozyryev2019determination}.

Briefly, we begin by constructing the Wilson force and kinetic energy-related matrices $\mathbf{F}$ and $\mathbf{G}$, which are related to the vibrational potential energy $V$ and kinetic energy $K$ by
\begin{equation}
2V = \left(\vec{S}\right)^T \mathbf{F} \vec{S}, \quad 2K = \left(\dot{\vec{S}}\right)^T \mathbf{G}^{-1} \dot{\vec{S}}
\end{equation}
where $\vec{S}$ is a vector of $3N-5$ internal molecular coordinates and $[\ldots]^T$ denotes the vector transpose. For CaOH, we define the internal coordinates so that $S_1 = \Delta r_{31}$, $S_2 = \Delta r_{32}$ and $S_3 = \left(r_{31} r_{32} \right)^{1/2} \Delta \phi$, where $r_{31}$ and $r_{32}$ are the equilibrium Ca--O and O--H bond lengths, respectively, $\Delta r_{31}$ and $\Delta r_{32}$ give the change in bond length, and $\Delta \phi$ is the change in the bending angle. From geometrical arguments the $\mathbf{G}$ matrix can be expressed as~\cite{wilson1955}
\begin{align}
\mathbf{G} = 
&\begin{pmatrix}
\mu_1 + \mu_3 & -\mu_3 & 0 \\
-\mu_3 & \mu_2 + \mu_3 & 0 \\
0 & 0 & G_{33}
\end{pmatrix} \text{ ,} \nonumber \\
& G_{33} = \mu_1 \frac{r_{32}}{r_{31}} + \mu_2 \frac{r_{31}}{r_{32}} + \mu_3 \frac{(r_{31} + r_{32})^2}{r_{31} r_{32}}
\end{align}
where $\mu_1 = 1/m_\text{Ca}$, $\mu_2 = 1/m_\text{H}$, $\mu_3 = 1/m_\text{O}$.

The $\mathbf{F}$ matrix consists of second derivatives of the potential energy surface with respect to the internal coordinates. Because there is at most only very weak coupling between the stretching and bending vibrations, we write $\mathbf{F}$ as
\begin{equation}
\mathbf{F} = 
\begin{pmatrix}
F_{11} & F_{12} & 0 \\
F_{21} & F_{22} & 0 \\
0 & 0 & F_{33}
\end{pmatrix}
\end{equation}
Diagonalizing the matrix product $\mathbf{GF}$ is equivalent to solving the secular equation, and its eigenvalues and eigenvectors are the frequencies and coordinates of the normal modes of vibration. In particular, properly-normalized eigenvectors of $\mathbf{GF}$ form the columns of the $\mathbf{L}$ matrix, which transforms normal coordinates $\vec{Q}$ into internal coordinates $\vec{S} = \mathbf{L} \vec{Q}$.

For the $\widetilde{\text{A}}\rightarrow \widetilde{\text{X}}$ transition, we use measured vibrational frequencies and bond lengths from Ref.~\cite{li1996dye} to solve for the force constants $F_{ij}$. This calculation makes use of equations given in Refs.~\cite{li1995bending,li1996dye}. For the $\widetilde{\text{B}}\rightarrow \widetilde{\text{X}}$ transition we use the measured bond lengths from Ref.~\cite{dick2006optical}, in conjunction with vibrational frequency calculations from Ref.~\cite{taylor2005electronic}. There is insufficient data on the $\widetilde{\text{B}}$ state to extract the off-diagonal force constants (these require data from CaOD as well), so we make the approximation that $F_{21} = F_{12} = 0$ for the $\widetilde{\text{B}}$ state. We find that this has little impact on the final result.

After solving for the normal modes using the $\mathbf{GF}$ matrix approach, harmonic overlap integrals may be factored as
\begin{equation}
\langle \overline{v'' | v'} \rangle =  \prod_i \int \psi_{v_i''}^* \psi_{v_i'} dQ_i \text{ ,}
\label{eqn:harmonicsep}
\end{equation}
where $\psi_{v_i}$ is a harmonic oscillator eigenfunction and $Q_i$ is a normal coordinate. Eq. \ref{eqn:harmonicsep} may be evaluated by transforming the normal coordinates $\vec{Q}''$ of the final state to those of the initial state $\vec{Q}'$ via a Duschinsky rotation,
\begin{equation}
\vec{Q}' = \mathbf{J} \vec{Q}'' + \vec{K}.
\end{equation}
The method employed here, due to Sharp and Rosenstock, relates $\mathbf{J}$ and $\vec{K}$ to the transformation matrices $\mathbf{L}''$, $\mathbf{L}'$ of the ground and excited states, as well as their equilibrium geometries. The solution makes use of generating functions, and the results may be calculated using standard computational tools~\cite{sharp1964franck, weber2003franck}.

In the case where the vibrational eigenstates are mixed by perturbations to the potential energy surface (as in the Renner-Teller and Fermi resonance interactions discussed above), FCFs may still be computed using the harmonic calculations described here. For example, if the ground and excited state vibrational wavefunctions are given by
\begin{equation*}
|v'\rangle = a |\overline{v_a'}\rangle + b | \overline{v_b'}\rangle \text{,} \quad |v''\rangle = c |\overline{v_a''}\rangle + d | \overline{v_b''}\rangle,
\end{equation*}
where the horizontal bar denotes a harmonic oscillator eigenstate, the FCF is
\begin{align*}
q_{v'-v''} =& \left| \langle v'' | v' \rangle \right|^2 \\
&= \left| a c^* \langle \overline{v_a'' | v_a'} \rangle + a d^* \langle \overline{v_a'' | v_b'} \rangle +  \ldots \right|^2
\end{align*}
where each term on the right-hand side is a harmonic overlap integral (Eq. \ref{eqn:harmonicsep}) and can be computed by the means described above.

\section{Details of Renner-Teller Calculations}
\label{app:RTHam}

The Renner Teller (RT) Hamiltonian may be written as ~\cite{brown1977effective,hirota2012high,bolman1973renner,northrup1990renner,presunka1994laser,li1997cacch}
\begin{equation}
    H_\text{RT} = \frac{V_{11}}{2}(q_+L_- + q_-L_+) + \frac{V_{22}}{2}(q_+^2L_-^2 + q_-^2L_+^2) + \ldots
\end{equation}
where $L_\pm$ and $q_\pm$ are ladder operators in the electronic angular momentum $\Lambda$ and the vibrational angular momentum $\ell$, respectively. The matrix elements of $q_\pm$ are \cite{Brown1983, li1997cacch, northrup1990renner}
\begin{align}
    &\langle v_2 + 1 , \ell \pm 1 \lvert q_\pm | v_2, \ell \rangle = \frac{1}{\sqrt{2}}(v_2 + 2 \pm \ell)^{1/2} \nonumber \\
    &\langle v_2 - 1 , \ell \pm 1 \lvert q_\pm | v_2, \ell \rangle = \frac{1}{\sqrt{2}}(v_2 \mp \ell)^{1/2}
\end{align}
while $L_\pm$ connects states satisfying $L_\pm|\Lambda\rangle \propto |\Lambda \pm 1\rangle$.
The first, dipolar, term of $H_\text{RT}$ therefore has matrix elements only between different electronic states, while the second, quadrupolar, term supports matrix elements within a single $\Pi$ state.

The direct vibronic mixing between $\widetilde{A}(000)$ and $\widetilde{B}(01^10)$ (Eq. 8 of the main text) may be derived using the definition (in the unique perturber approximation) of the spectroscopic parameter $g_K$  \cite{brown1977effective, northrup1990renner}:
\begin{equation}
    g_K \approx \frac{\omega_2 \lvert \langle \widetilde{B} \lvert V_{11} L_+ \rvert \widetilde{A} \rangle \rvert^2}{4\Delta E_{AB}^2}
\end{equation}
This parameter appears as an energy offset $g_K(\Lambda + \ell)$ in the effective Hamiltonian for vibronically perturbed levels of a $\Pi$ electronic state, and was previously fit for CaOH \cite{li1995bending, li1996dye}. The above definition may be used to evaluate
\begin{align}
    \langle \widetilde{B}(01^10)^2\Pi\lvert \widetilde{A}(000)^2\Pi_{1/2} \rangle &\approx \frac{\langle \widetilde{B}\lvert V_{11} L_+ | \widetilde{A} \rangle/2}{\Delta E_{AB} - \omega_2} \nonumber \\
\implies  \lvert \langle \widetilde{B}(01^10)^2\Pi \lvert \widetilde{A}(000)^2\Pi_{1/2} \rangle \rvert^2 &\approx \frac{g_K}{\omega_2\left(1-\frac{\omega_2}{\Delta E_{AB}}\right)^2} , \nonumber
\end{align}
which is Eq. 8 of the main text.

We now consider the $H_\text{RT} \times H_\text{SO}$ perturbation in more detail. The spin-orbit Hamiltonian is
\begin{equation}
    H_\text{SO} = A_\text{SO}L_z S_z +\frac{1}{2}A_\text{SO}(L_+S_- + L_-S_+)
\end{equation}
where $S_z, S_\pm$ are angular momentum operators acting on the electronic spin state $\lvert S,\Sigma\rangle$. Their matrix elements are defined as usual,
\begin{equation}
    \langle S, \Sigma \pm 1 \lvert S_\pm \rvert S, \Sigma \rangle = \left[ S(S+1) - \Sigma(\Sigma \pm 1) \right]^2
\end{equation}

The $\widetilde{A}^2\Pi(010)$ state is split into several vibronic components due to Renner-Teller and spin-orbit interactions, as shown in Fig. \ref{fig:A010mixing} of App. \ref{app:structure} \cite{hougen1962rotational, li1995bending}. Both the $\widetilde{A}(010)\mu^2\Sigma^{(+)}$ and $\widetilde{A}(010)\kappa^2\Sigma^{(-)}$ components are mixed with $\widetilde{A}(000)^2\Pi_{1/2}$ and contribute to vibronic decay. The second-order mixing amplitudes of these states due to the $H_\text{RT} \times H_\text{SO}$ perturbation are \cite{li1997cacch}
\begin{align}
    \langle \widetilde{A}(01&0)\mu^2\Sigma^{(+)}\lvert \widetilde{A}(000)^2\Pi_{1/2} \rangle \nonumber \\
    &\approx 2\frac{\langle \widetilde{A} \lvert A_\text{SO} L_- \rvert \widetilde{B}\rangle \langle \widetilde{B} \lvert V_{11} L_+ \rvert \widetilde{A}\rangle}{4\omega_2 \Delta E_{AB}}
    \times (\cos \beta - \sin \beta) \label{eqn:muRT}\\
    \langle \widetilde{A}(01&0)\kappa^2\Sigma^{(+)}\lvert \widetilde{A}(000)^2\Pi_{1/2} \rangle \nonumber \\
    &\approx 2\frac{\langle \widetilde{A} \lvert A_\text{SO} L_- \rvert \widetilde{B}\rangle \langle \widetilde{B} \lvert V_{11} L_+ \rvert \widetilde{A}\rangle}{4\omega_2 \Delta E_{AB}}
    \times (\cos \beta + \sin \beta) \label{eqn:kappaRT}
\end{align}
where the prefactor of 2 comes from the two contributions via $\widetilde{B}(000)$ and $\widetilde{B}(01^10)$, respectively. The mixing angle $\beta$ defines the vibronic character of the $\mu$ and $\kappa$ states \cite{hougen1962rotational, hirota2012high},
\begin{equation}
    \beta = \frac{1}{2}\arcsin \left(\frac{2 \epsilon \omega_2}{\sqrt{(2 \epsilon \omega_2)^2+A_\text{SO}^2}} \right)
    \label{eqn:RTmixingangle}
\end{equation}
where $\epsilon$ is the Renner parameter.

The common factor in eqns. \ref{eqn:muRT}-\ref{eqn:kappaRT} may be evaluated using the pure precession approximation, $\langle \widetilde{A}|\frac{1}{2}A_\text{SO} L_-|\widetilde{B}\rangle \approx A_\text{SO}/\sqrt{2}$, which holds well in CaOH \cite{hilborn1983laser, bernath1984dye}. The result is
\begin{equation}
    \left| \frac{\langle \widetilde{A} \lvert A_\text{SO} L_- \rvert \widetilde{B}\rangle \langle \widetilde{B} \lvert V_{11} L_+ \rvert \widetilde{A}\rangle}{2\omega_2 \Delta E_{AB}} \right|^2 \approx \frac{2 g_K A_\text{SO}^2}{\omega_2^3}
\end{equation}
which may be used to find the mixing probabilities,
\begin{align}
    | \langle \widetilde{A}(01&0)\mu^2\Sigma^{(+)}\lvert \widetilde{A}(000)^2\Pi_{1/2} \rangle |^2 \approx 1.9\times 10^{-4} \nonumber \\
    | \langle \widetilde{A}(01&0)\kappa^2\Sigma^{(+)}\lvert \widetilde{A}(000)^2\Pi_{1/2} \rangle |^2 \approx 3\times 10^{-5}
\end{align}
The sum of these contributions is double the mixing fraction that would be calculated if the vibronic structure of the $\widetilde{A}^2\Pi(01^10)$ manifold were ignored.

The Renner parameter used to quantify $\Delta \ell = \pm 2$ mixing is $\epsilon \approx \epsilon^{(1)} + \epsilon^{(2)}$, where the first-order contribution,
\begin{equation}
    \epsilon^{(1)} \approx \langle \tilde{A} \lvert V_{22} L_-^2 \rvert \tilde{A} \rangle / \omega_2
\end{equation}
quantifies direct quadrupolar mixing within the $\widetilde{A}$ state. The second-order contribution is
\begin{equation}
    \epsilon^{(2)} \approx - \frac{\lvert \langle \widetilde{A} \lvert V_{11} L_-\rvert \widetilde{B} \rangle \rvert ^2}{2 \omega_2 \Delta E_{AB}} \approx - \frac{2 g_K \Delta E_{AB}}{\omega_2^2} , 
\end{equation}
and arises due to dipolar mixing via the $\widetilde{B}$ state \cite{brown1977effective, hirota2012high}.

\section{Fermi Resonance Matrix Elements}
\label{app:Fermi}

The nonzero matrix elements of the Fermi resonance operator $V_F = k_{122}Q_1 Q_2^2$ for a non-degenerate vibronic state are~\cite{hougen1962fermi}
\begin{subequations}
\begin{align}
    \langle v_1+1,v_2,v_3;l|V_F|&v_1,v_2+2,v_3;l\rangle \nonumber \\
    = W \big[(v_1&+1)(v_2+2-l)(v_2+2+l)\big]^{1/2} \\
    \langle v_1+1,v_2,v_3;l|V_F|&v_1,v_2,v_3;l\rangle \nonumber \\
    &= 2W (v_1+1)^{1/2}(v_2+1)
\label{eqn:Wmatrixelements}
\end{align}
\end{subequations}
where
\begin{equation}
W = \frac{k_{122}}{2\sqrt{2}}\left(\frac{\hbar}{2\pi c \omega_2} \right)\left(\frac{\hbar}{2\pi c \omega_1} \right)^{1/2}
\label{eqn:FermiW}
\end{equation}
is the Fermi resonance parameter and depends on only the force constant $k_{122}$, the Ca--O stretching frequency $\omega_1$, and the Ca--O--H bending frequency $\omega_2$. Here $c$ is the speed of light and $\omega_{1,2}$ have units of cm$^{-1}$. Because the second matrix element (Eq. \ref{eqn:Wmatrixelements}) connects states separated by a relatively large energy, its effects are neglected in this work. We note, however, that matrix elements of this sort generically arise from other anharmonic terms in the potential energy function as well. These couplings may contribute to higher-order decay channels, though a full analysis of such effects is beyond the scope of this work.

While the Fermi resonance parameter $W$ has not been measured for the $\widetilde{\text{X}}\,^2\Sigma^+$ state of CaOH, it has been fit in the $\widetilde{\text{A}}\,^2\Pi$ state from an analysis of the $(100)\sim(020)$ Fermi dyad~\cite{li1996dye}. Assuming that $k_{122}$ is unchanged between the $\widetilde{\text{X}}$ and $\widetilde{\text{A}}$ states, and using measured vibrational frequencies $\omega_{1,2}$~\cite{li1996dye}, we estimate a Fermi resonance parameter $|W_{X}| \approx 10.7$ cm$^{-1}$ in the $\widetilde{\text{X}}\,^2\Sigma^+$ state by scaling the $\widetilde{\text{A}}$ state measurement by the appropriate factors in Eq. \ref{eqn:FermiW}. This is in good agreement with a separate estimate $|W_{X}| \approx 11.1$ cm$^{-1}$, found using a deperturbative analysis (similar to one in Ref. \cite{li1996dye}) of the experimentally observed splittings between the $\widetilde{\text{X}}(02^00)-\widetilde{\text{X}}(02^20)$ and $\widetilde{\text{X}}(12^00)-\widetilde{\text{X}}(12^20)$ (Ref. \cite{coxon1992investigation} and Appendix C of the main text). While the sign of $W$ was not determined experimentally in Ref.~\cite{li1996dye}, by relating $W$ to other measured constants it was deduced in that work that $W < 0$.

\bibliographystyle{apsrev4-1}
\bibliography{CaOHStructure_References_v0}

\end{document}